\documentclass{jkas}

\def\year{2022} 
\def\month{October} 
\def\volume{55} 
\def\issue{5} 
\def\beginpage{149} 
\def\endpage{172} 
\def\received{June 30, 2022} 
\def\accepted{August 29, 2022} 
\date{Received \received; accepted \accepted}

\newcommand\ion[2]{{#1}\,{\sc #2}} 
\newcommand\arcsec{$^{\prime\prime}$}
\newcommand\kms{km\,s$^{-1}$}
\newcommand\an{$A_{\rm{n}}$}
\newcommand\ab{$A_{\rm{b}}$}
\newcommand\at{$A_{\rm{tot}}$}
\newcommand\snsb{$\sigma_{\rm{n}} / \sigma_{\rm{b}}$}
\newcommand\anab{$A_{\rm{n}} / A_{\rm{b}}$}
\newcommand\anat{$A_{\rm{n}} / A_{\rm{tot}}$}

\newcommand\HI{\ion{H}{i}\xspace}

\usepackage[colorlinks,
            breaklinks,
            linkcolor=blue,
            urlcolor=blue,
            anchorcolor=blue,
            menucolor=blue,
            citecolor=blue]{hyperref}

\usepackage{amsmath, graphicx, natbib}
\usepackage{array,booktabs,multirow, makecell}

\usepackage{adjustbox}
\usepackage{xspace}
\usepackage{tablefootnote}
\usepackage{longtable}
\usepackage{booktabs}
\usepackage{graphicx}
\usepackage{float}
\usepackage{subfig}

\newcommand{\pt}[1]{\phantom{#1}}

\newcolumntype{x}[1]{>{\raggedright\arraybackslash}m{#1}}
\newcolumntype{y}[1]{>{\centering\arraybackslash}m{#1}}
\newcolumntype{z}[1]{>{\raggedleft\arraybackslash}m{#1}}
\title{
Global \ion{H}{i} Properties of Galaxies via Super-Profile Analysis
}

\author[1]{\href{https://orcid.org/0000-0002-1776-6526}{Minsu Kim}}
\author[1,2]{\href{https://orcid.org/0000-0002-8379-0604}{Se-Heon Oh}}

\affil[1]{Department of Astronomy and Space Science, Sejong University, Seoul~05006, Korea}
\affil[2]{Department of Physics and Astronomy, Sejong University, Seoul~05006, Korea; \email{seheon.oh@sejong.ac.kr}}

\def\corrauthor{
S.-H. Oh
}

\def\runningauthor{
Kim \& Oh
}

\def\runningtitle{
Global \ion{H}{i} Properties of Galaxies via Super-Profile Analysis
}

\def\keywords{
galaxies: dwarf --- galaxies: ISM --- galaxies: irregular --- galaxies: kinematics and dynamics --- ISM: kinematics and dynamics
}

\def\abstracttext{
We present a new method which constructs an \HI super-profile of a galaxy which is based on profile decomposition analysis. The decomposed velocity profiles of an \HI data cube with an optimal number of Gaussian components are co-added after being aligned in velocity with respect to their centroid velocities. This is compared to the previous approach where no prior profile decomposition is made for the velocity profiles being stacked. The S/N improved super-profile is useful for deriving the galaxy’s global \HI properties like velocity dispersion and mass from observations which do not provide sufficient surface brightness sensitivity for the galaxy. As a practical test, we apply our new method to 64 high-resolution \HI data cubes of nearby galaxies in the local Universe which are taken from THINGS and LITTLE THINGS. In addition, we also construct two additional \HI super-profiles of the sample galaxies using symmetric and all velocity profiles of the cubes whose centroid velocities are determined from Hermite $h_3$ polynomial fitting, respectively. We find that the \HI super-profiles constructed using the new method have narrower cores and broader wings in shape than the other two super-profiles. This is mainly due to the effect of either asymmetric velocity profiles’ central velocity bias or the removal of asymmetric velocity profiles in the previous methods on the resulting \HI super-profiles. We discuss how the shapes (\snsb, \anab, and \anat) of the new \HI super-profiles which are measured from a double Gaussian fit are correlated with star formation rates of the sample galaxies and are compared with those of the other two super-profiles.
}

\begin{document}
\jkashead 


\section{Introduction\label{sec:s1}}

Deriving physical properties of neutral hydrogen (\HI) gas in galaxies is important for understanding galaxy formation and evolution. Not only does \HI acts as a gas reservoir for star formation but also it is a useful probe for tracing hydrodynamical processes in galaxies such as stellar feedback (e.g., \citealt{Heiles79, Boomsma+08, Hony+15, Bacchini+20}), gas kinematics (e.g., \citealt{Bosma78, Binney92, Walter+08, Oh+15}), and galaxy environmental effects (e.g., \citealt{Clemens+00, Vollmer+04, Zwaan+05}). As a kinematic tracer of galaxies, \HI particularly benefits from its extended distribution in the gas disk, usually well beyond the stellar disk. This makes it an excellent tool for studying galaxy kinematics and mass distribution to the very outskirts. 

One of the key observational properties of galaxies that can be obtained from \HI observations is the \HI gas velocity dispersion. The interstellar medium (ISM) can be distrubed by hydrodynamic and gravitational forces from baryonic processes in galaxies such as star formation (e.g., for review, see \citealt{Krumholz+14}) and tidal interactions in and around galaxies (e.g., \citealt{Agertz+09}; \citealt{Bournaud+11}; \citealt{Renaud+14}; \citealt{Dale+19}). These often give rise to turbulent gas motions in galaxies, either locally or globally, which tend to have higher gas velocity dispersion, and/or even deviate from the underlying kinematics of the galaxies (see, e.g., \citealt{Tamburro+09}; \citealt{Bacchini+20}). Therefore, the measurement of gas velocity dispersions of galaxies via \HI observations is needed to examine the interplay between the ISM and hydrodynamical processes. 

\HI gas velocity dispersion of a galaxy is prone to be under- or over-estimated, particularly in the central regions when using its global profile derived from low angular resolution single dish observations. If the spatial resolution is not high enough (i.e., a large beam size) to resolve the internal gas disk of the galaxy, multiple gas clouds at different line-of-sight velocities can be co-located within the beam. This usually results in higher gas velocity dispersions. This so-called beam smearing effect is the most severe toward the central region of a galaxy where the gradient of line-of-sight velocities increases. This effect would be evident for a galaxy whose angular size is smaller than a telescope's beam size (\citealt{Davies+11}).

On the other hand, high angular resolution observations with radio interferometries less suffer from the beam smearing effect on the measurement of \HI gas velocity dispersions of galaxies. They are able to provide spatially resolved information about the gas velocity dispersion of their gas disks. For example, sub-kpc resolution \HI data from THINGS\footnote{The \HI Nearby Galaxy Survey} \citep{Walter+08} and LITTLE THINGS\footnote{Local Irregulars That Trace Luminosity Extremes, The \HI Nearby Galaxy Survey} \citep{Hunter+12} were used for examining the relationship between spatially resolved \HI gas velocity dispersions and star formation rates in dwarf and spiral galaxies (\citealt{Warren+12}; \citealt{Cigan+16}; \citealt{Ashley+17}; \citealt{Krumholz+18}).

However, this can be only achieved for the galaxies whose gas disks are spatially well-resolved by the beams of radio interferometries (e.g., $\geq$6 beams along the semi-major axes of galaxies; \citealt{deBlok+08}, \citealt{Oh+11, Oh+15}). Even with the currently available radio interferometries like the VLA\footnote{NRAO Very Large Array}, WSRT\footnote{Westerbork Synthesis Radio Telescope}, GMRT\footnote{Giant Metrewave Radio Telescope}, and ATCA\footnote{Australia Telescope Compact Array} which have $\sim$10\arcsec\ beam resolutions, sub-kpc resolution observations at \HI are only achievable for the galaxies within a redshift $z$ of $\sim$0.005 ($\sim$20~Mpc). This situation will continue even in the upcoming Square Kilometre Array (SKA) era as its observational parameter space will be prefentially extended towards the low column density levels \citep{Dewdney+09}. More fundamentally, both high-resolution interferometry and single dish \HI observations are not free from an inevitable decrease of S/N with increasing distances to target galaxies. Low \HI column density features are particularly prominent in the outer regions of galaxies where an \HI break is expected by photoionization (e.g., \citealt{Maloney93}, \citealt{Dove&Shull94}).

\defcitealias{Ianja+12}{I12}
Stacking of velocity profiles of an \HI data cube of a resolved galaxy is able to provide robust measurements of the \HI gas velocity dispersion. This can improve the low S/N issue at the cost of loss of spatial information. \cite{Ianja+12} (hereafter \citetalias{Ianja+12}) carried out stacking of \HI velocity profiles of THINGS spiral galaxies to derive their global \HI properties. The so-called \HI super-profile is constructed by co-adding individual line profiles after aligning them in velocity with their centroid velocities. The S/N of the stacked profile scales with $\sqrt{N}$ as the noise increases with $\sqrt{N}$, where $N$ is the number of independent velocity profiles being stacked.

\citetalias{Ianja+12} decomposed an \HI super-profile into kinematically narrower (smaller velocity dispersion) and broader (larger velocity dispersion) Gaussian components by fitting a double Gaussian model and estimated their velocity dispersions. On the other hand, \cite{Stilp+13} parameterized an \HI super-profile in a way of matching the profile's peak flux and full-width-half-maximum (FWHM) with a single Gaussian function. Flux residuals, if present, in the profile's wing regions can be attributed to deviating gas motions from the galaxy's global kinematics. As discussed in \citetalias{Ianja+12} and \cite{Stilp+13}, these velocity dispersion measurements derived from the \HI super-profile analysis can be correlated with other physical properties of galaxies like metallicity, FUV-NUV colors, star formation rate (SFR), H$\alpha$ luminosities, halo mass, etc.

Previous stacking methods use the central velocities of profiles which are determined from the moment analysis \citepalias{Ianja+12} or the analysis of alternative fitting forms such as single Gaussian \citep{Mogotsi+16} and Gauss-Hermite polynomial functions (\citealt{Ianja+12, Ianja+15}; \citealt{Stilp+13}; \citealt{Faridani+14}; \citealt{Patra20}; \citealt{Saponara+20}; \citealt{Yadav+21}; \citealt{Hunter+22}).
Turbulent gas motions caused by hydrodynamical processes in galaxies combined with the beam smearing effect often make their velocity profile shape non-Gaussian and asymmetric. The centroid velocities derived using the conventional methods can be biased. Therefore, a super-profile constructed using these asymmetric non-Gaussian velocity profiles will have broader (and even asymmetric) wings while having lower peak flux than the ones for the galaxies that are less affected by turbulent gas motions. The resulting velocity dispersion and peak flux of the super-profile are over- and under-estimated, respectively.

To minimize the effect of turbulent random gas motions in galaxies and observational beam smearing on their stacked profiles, we present a new method that constructs an \HI super-profile of a resolved galaxy via profile decomposition based on Bayesian analysis. This method performs profile decomposition of individual line profiles of an input data cube based on Bayesian nested sampling techniques. From this, each line profile is decomposed into an optimal number of Gaussian components. To this end, we use a tool, {\sc baygaud}\footnote{\url{https://github.com/seheonoh/baygaud}} which performs multiple Gaussian decomposition of line profiles via Bayesian model selection and parameter estimation (\citealt{Oh+19}). We align the optimally decomposed Gaussian components with respect to their central velocities determined and co-add them to construct a super-profile. This new method is capable of minimizing the effect of turbulent gas motions and beam smearing in galaxies on the super-profile. In this work, we present the details of the new method and its practical application to the high-resolution \HI data cubes of nearby galaxies from THINGS and LITTLE THINGS.

This paper is arranged as follows. In Section~\ref{sec:s2}, we describe our method of constructing \HI super-profiles. We then make a comparison with the conventional stacking methods with ours in Section~\ref{sec:s3}. In Section~\ref{sec:s4}, we investigate correlations between the parameters of super-profiles and physical properties of galaxies for a practical test of our stacking method. Lastly, in Section~\ref{sec:s5}, we summarize the main results.

\begin{figure*}[!t]
\centering
\includegraphics[width=\textwidth, trim={0mm, 2mm, 0mm, 0mm}, clip]{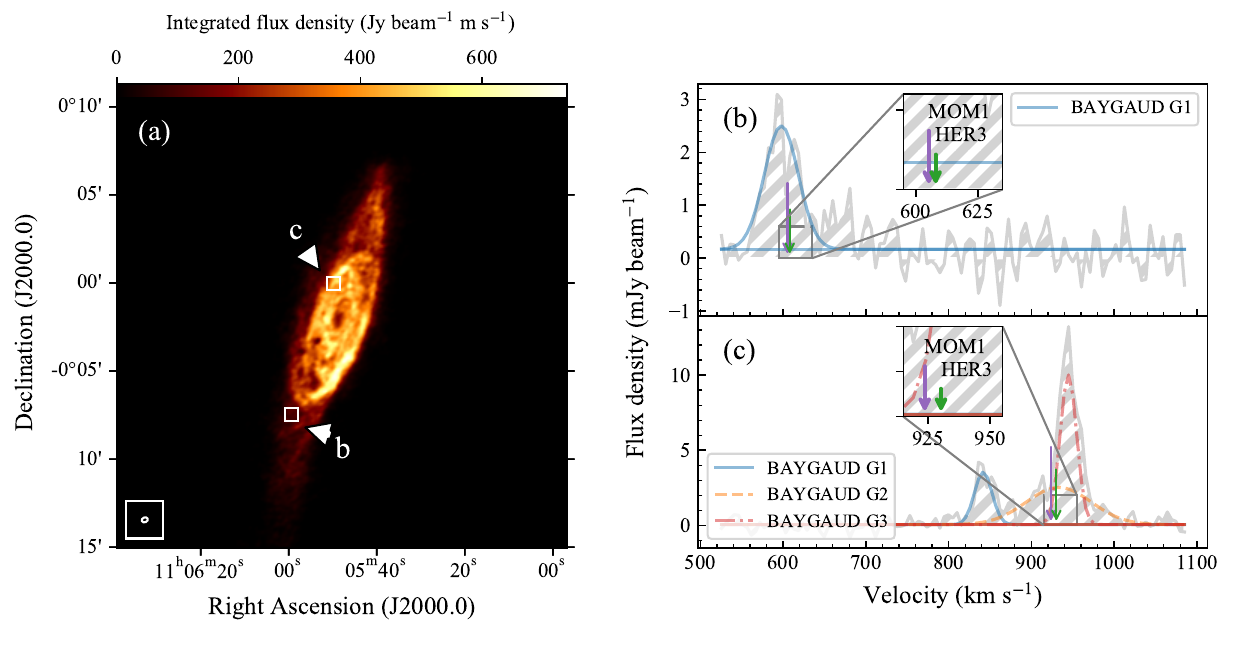}
\caption{An example of the profile analysis using {\sc baygaud}. Two examples of {\sc baygaud} fit result on \HI velocity profile are shown along with conventional centroid velocity estimates ({\sc moment1} and Gauss-Hermite $h_3$ polynomial). Panel (a): the \HI integrated intensity map ({\sc moment0}) of NGC\,3521 from THINGS, (b): an example \HI velocity profile in the outskirt, and (c): an example \HI velocity profile in the disk region. The profile on panel (b) is fitted with a single Gaussian component while the profile in panel (c) is fitted with triple Gaussian components. The purple and green arrows show the centroid velocity estimates from {\sc moment1} and Gauss-Hermite $h_3$ polynomial, respectively. The white circle on the bottom left of the panel (a) is the beam size of the observation.\label{fig:fig1}}
\vspace{2mm}
\end{figure*}

\section{Hi Super-Profiles\label{sec:s2}}

In this section, we describe the procedures of our stacking method for constructing an \HI super-profile of a galaxy. An input 3-dimensional (right ascension; R.A., declination; Dec., and velocity) data cube of \HI spectra whose angular and velocity resolutions are high enough to resolve a galaxy both spatially and spectrally is a prerequisite. Estimation of global shape parameters of the super-profile and their uncertainties constructed are also discussed.

\subsection{Profile Decomposition\label{sec:s21}}

As the first step of constructing an \HI super-profile of a galaxy, we model individual velocity profiles of the input data cube with a set of multiple Gaussian components. Multiple kinematic components of \HI gas are often present along a line-of-sight in the gas disk of a galaxy, which can be grouped into two kinematic populations: 1) bulk gas motions rotating with the underlying global galaxy kinematics, and 2) non-circular random (or streaming) gas motions deviating from the global kinematics. This makes the line-of-sight gas velocity profile non-Gaussian and/or asymmetric. It is common to observe non-Gaussian asymmetric velocity profiles from high-resolution \HI observations of galaxies, particularly for those undergoing significant hydrodynamical and/or gravitational processes. We, therefore, need to de-blend the velocity profile into bulk and random gas motions to better understand the coupled kinematics.

\begin{figure*}[!t]
\centering
\includegraphics[width=\linewidth, trim={0mm, 5mm, 0mm, 5mm}, clip]{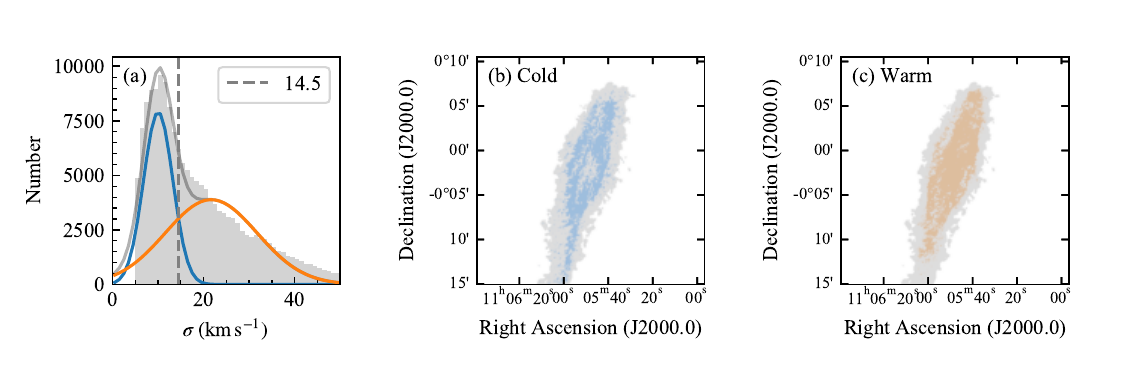}
\caption{The spatial distribution of cold and warm components of NGC\,3521. (a): a distribution of velocity dispersions overplotted with the double Gaussian model fit, and (b, c): spatial distribution of cold and warm components, respectively.} \label{fig:distribution_coldwarm}
\vspace{2mm}
\end{figure*}

To this end, we use a new tool {\sc baygaud}, which allows us to decompose a line-of-sight \HI velocity profile into an optimal number of Gaussian components based on Bayesian analysis techniques. It fits a series of models with a different number of Gaussian components to each velocity profile and finds the most appropriate one via Bayesian model selection. For the model selection, it computes Bayes factors between any two competing models on the trials and finds the best model whose Bayesian evidence is at least larger than 10 times the one of the second-best model (e.g., `strong' model selection criteria). From this, the input velocity profile can be parameterized with an optimal set of Gaussian components. The best fit values of individual Gaussian parameters (i.e., peak flux, centroid velocity, and velocity dispersion) and their uncertainties are derived together. We refer to \cite{Oh+19} for more details of the profile decomposition analysis.

Figure~\ref{fig:fig1} shows an example of the profile analysis using {\sc baygaud}. Panel (a) shows the \HI integrated intensity map ({\sc moment0}) of NGC\,3521 taken from THINGS observations. Two example \HI velocity profiles in the outskirt and disk regions of the galaxy are presented in the panels (b) and (c), respectively. From the profile analysis of the profiles using {\sc baygaud}, the profile taken from the outskirt region is reasonably described by a single Gaussian function. The derived centroid velocity is consistent with the one derived from the moment analysis ({\sc mom1}: purple arrow) and Hermite $h_3$ fit ({\sc her3}: green arrow). On the other hand, the one in the disk region which shows a non-Gaussian and asymmetric feature in profile shape is better described by a triple Gaussian model. The moment analysis as well as the Hermite $h_3$ polynomial fitting do not fully account for the asymmetric non-Gaussian feature of the profile. The derived centroid velocities using both the Hermite $h_3$ polynomial fitting and moment analysis are significantly different from the one from the {\sc baygaud} analysis. Evidently, the kinematics of the corresponding gas cloud is not well explained by a single representative centroid velocity. In this way, we decompose all the line-of-sight velocity profiles of the input \HI data cube into an optimal number of Gaussian components and parameterize them.

Additionally, {\sc baygaud} can also be used to locate the cold and warm \HI components inside a galaxy. In Figure~\ref{fig:distribution_coldwarm}, we show this by using NGC\,3521 as an example, following the method described in \cite{Park+22}. Panel (a) shows the distribution of {\sc baygaud} velocity dispersions of the decomposed Gaussian component in the cube. We fit a two-Gaussian model to the histogram as it shows a bimodality. From this we derive a velocity dispersion limit of 14.5~\kms, which the cold and warm components are separated. Panels \mbox{(b, c)} show the 2d maps of the cold and warm components separated, respectively.

\begin{figure}[!t]
\centering
\includegraphics[width=\linewidth, trim={0mm 5mm 0mm 5mm}, clip]{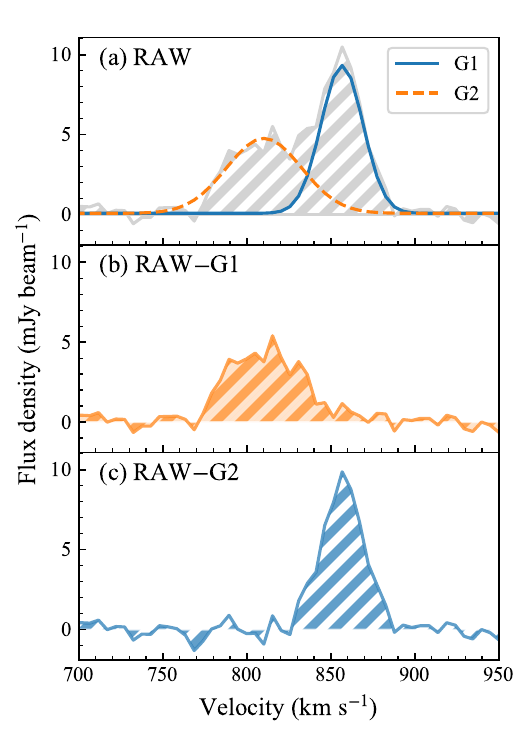}
\caption{The component isolating procedure is shown with an example velocity profile decomposed with two Gaussian components: G1 and G2. Panel (a): a raw velocity profile, (b): an isolated profile by subtracting model G1 from the raw profile, and (c): an isolated profile by subtracting model G2 from the raw profile.\label{fig:2Gisolate}}
\vspace{2mm}
\end{figure}

\subsection{Stacking of the Decomposed Velocity Profiles\label{sec:s22}}

We stack the decomposed velocity profiles of the input data cube after aligning them in velocity with their centroid velocities which are determined as described in Section~\ref{sec:s21}. Unlike the line-of-sight velocity profile which is best described by a single Gaussian component, a profile modeled with multiple Gaussian components needs to be separated before being stacked together. To isolate one kinematic component from multithe ple Gaussian components in a velocity profile, we subtract the flux contribution of the additional kinematic components from the total flux of the profile. That is, the sum of fluxes that are modeled by the additional Gaussian models along the velocity axis is subtracted from the input raw velocity profile. The residual fluxes along the velocity axis are attributed to the kinematic component we are interested in. The rest of the other kinematic components in the velocity profile is isolated in the same way. Figure~\ref{fig:2Gisolate} shows this process with an example velocity profile fitted with double Gaussian components.

The velocity profiles with low S/N (e.g., $<$3) are omitted in the stacking process. This is for minimizing the effect of uncertainties in the centroid velocity measurements of low S/N profiles on the superprofile. The misaligned low S/N velocity profiles in velocity will result in broad wings in the super-profile. In addition, we note that when stacking the decomposed velocity profiles we do not use the Gaussian model profiles but we use the residual ones from which the {\it Gaussian model profiles} for the other kinematic components are subtracted (see Figure~\ref{fig:2Gisolate}). This is for taking the effect of noise characteristics of the velocity profiles on the resulting super-profile into account. We use the centroid velocities of the model Gaussian profiles when aligning their residual velocity profiles. For a velocity profile best modeled by a single Gaussian function, we directly stack it using the centroid velocity derived from the fitting without any pre-processing. Lastly, the velocity profiles corresponding to the optimally decomposed Gaussian components are background subtracted, aligned with respect to their centroid velocities, and co-added in order to construct an \HI velocity profile.

\begin{figure}[!t]
\centering
\includegraphics[width=\linewidth, trim={0mm, 3mm, 0mm, 3mm}, clip]{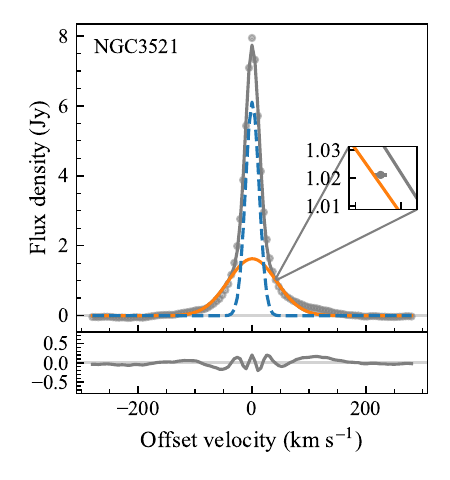}
\caption{An example of {\sc 2gfit} parameterization of a super-profile built from NGC\,3521 of THINGS. The plotted error bars show 3-$\sigma$ uncertainties for a given point. The bottom panel shows residuals of the fit.\label{fig:2GFIT_NGC3521}}
\vspace{2mm}
\end{figure}

\subsection{Parameterizing the Shapes of Super-Profiles\label{sec:s23}}

We parameterize the global shape of an \HI super-profile constructed in the previous section by fitting a double Gaussian model. There are possible model variants for parameterizing the super-profile which, for example, include a Gaussian model with more than two components with or without having non-zero centroid velocities. In this work, as a demonstration, we use a double Gaussian model of which individual Gaussian components do not necessarily have the same peak flux and velocity dispersion values but share an identical centroid velocity. \citetalias{Ianja+12} also adopt the same parameterization for the \HI super-profiles (but \citealt{Stilp+13} parameterize an \HI super-profile by scaling a single Gaussian function in a way of matching the amplitude and FWHM of the function to the peak flux and FHWM of the superprofile). Hereafter, we call the super-profile parameterization method adopted in this work as {\sc 2gfit}.

In Figure~\ref{fig:2GFIT_NGC3521}, we show an example of the {\sc 2gfit} to an \HI super-profile of NGC\,3521 constructed in this work. The decomposed two Gaussian components from the \HI super-profile are classified as kinematically narrow (blue dotted line) and broad (orange solid line) ones which have lower and higher velocity dispersions, respectively. As discussed earlier, gas velocity dispersions are associated with the kinetic energy of gas clouds in the host galaxy which is deposited by hydrodynamical and/or gravitational processes. This will be further discussed later in Section~\ref{sec:s32}.

We use a python package, \texttt{emcee}\footnote{\url{https://github.com/dfm/emcee}} to perform the {\sc 2gfit} parameterization for an \HI super-profile. \texttt{emcee} is capable of fitting a nonlinear model to the input data using Markov Chain Monte Carlo (MCMC) techniques. We refer to the \texttt{emcee} webpage for the detailed fitting algorithm and performance. The parameter setup for the {\sc 2gfit} analysis using \texttt{emcee} with six free parameters is as follows:

\begin{itemize}
\small
\item B: the constant baseline of the super-profile
\item $a_1$: the amplitude of the first Gaussian component
\item $\sigma_1$: the velocity dispersion of the first Gaussian component
\item $a_2$: the amplitude of the second Gaussian component
\item $\sigma_2$: the velocity dispersion of the second Gaussian component
\item $v$: the central velocity of the Gaussian components .
\end{itemize}

\noindent Here, in contrast to the {\sc 2gfit} parameterization in \citetalias{Ianja+12}, we set the central velocities to be the same for the first and second Gaussian components. Then, following \citetalias{Ianja+12}, we quantify the shape of a super-profile by estimating 1) \snsb, 2) \anab, and 3) \anat\ where the subscipts n and b indicate the Gaussian parameters with the narrower and broader velocity dispersion between the two Gaussians. The integrated intensity of a Gaussian component $A$ is estimated using its corresponding $\sigma$ and $a$ (i.e., $\sqrt{2\pi}\sigma a$). \at\ is the total total area of the super-profile which equals to \an+\ab.

\subsection{Fitting Weights of {\sc 2gfit} Analysis}
Following \citetalias{Ianja+12}, we estimate the flux uncertainty in each data point of the super-profile at a velocity channel as follows,
\begin{equation}
\sigma = \sigma_{\rm{chan}} \times \sqrt{N / N_{\rm{prof}}^{\rm{beam}}}
\end{equation}
\noindent where $\sigma_{\rm{chan}}$ is the rms noise per channel of the input data cube, $N$ is the total number of data points which are co-added at the channel, and $N_{\rm{prof}}^{\rm{beam}}$ is the number of pixels per beam of the cube.
We use the same channel noise, $\sigma_{\rm{chan}}$ throughout the cube. There is a possibility of non-uniform noise levels through the cube. This can be mainly caused by the primary beam correction in the course of data calibration. However, as discussed in \cite{deBlok&Walter+06}, in most cases it would be minimal as the area of a galaxy usually occupies a fraction of the primary beam. When carrying out the {\sc 2gfit} analysis on super-profiles, we use $1/\sigma^{2}$ as a fitting weight. Note that the error bars shown in Figure~\ref{fig:2GFIT_NGC3521} indicate 3-$\sigma$ uncertainties which are much smaller than the symbol size in most cases as in \citetalias{Ianja+12}.


\subsection{Estimating Uncertainties for Super-Profile Parameters}
To estimate the uncertainties of super-profile parameters, we perform a resampling of the parameters by adding random noise to the super-profile. We derive the rms of the flux residuals from the {\sc 2gfit} analysis, and use it as the standard deviation of the normal distribution for the random flux noise. We fix the background level of the super-profile, and perform a large number of iterations (e.g., $>$500) of the resampling using the python package \texttt{scipy.optimize.minimize}. We estimate the {\sc 2gfit} parameters of the generated super-profiles, and derive their standard deviations which are adopted as uncertainties of the corresponding parameters.

\section{Comparison with Previous Methods\label{sec:s3}}
In this section, we compare characteristics of our \HI super-profile method with those of the method described in \citetalias{Ianja+12} which parameterizes an \HI super-profile by fitting a double Gaussian model. As described in Section~\ref{sec:s2}, the key difference between the two methods is in the treatment of a line-of-sight velocity profile with a single \citepalias{Ianja+12} or multiple kinematic (ours) components. Additionally, \citetalias{Ianja+12} estimate the centroid velocity of a line profile using Hermite $h_3$ function (hereafter {\sc her3}), and our method derives it by fitting a single Gaussian function to each kinematic component decomposed.

\subsection{Profile Analysis of the H\textsc{i} Data Cube of NGC 3521\label{sec:s31}}

\begin{table}[!t]
\centering
\caption{Basic physical properties of NGC\,3521.\label{tab:ngc3521}}
\begin{tabular}{x{0.6\linewidth} y{0.4\linewidth}}
\toprule
\quad Right ascension [J2000.0]               & \quad \,11$^{\rm{h}}$05$^{\rm{m}}$48$^{\rm{s}}.$6\rlap{$^{\rm{a}}$}\\
\quad Declination [J2000.0]                   & $-$10$^{\circ}$02$'$09\arcsec.2\rlap{$^{\rm{a}}$}\\
\quad Distance [Mpc]                          & 10.7\rlap{$^{\rm{a}}$}\\
\quad Systemic velocity $V_{\rm{sys}}$ [\kms] & 803.5\rlap{$^{\rm{b}}$} \\
\quad Inclination [$^{\circ}$]                & 72.7\rlap{$^{\rm{b}}$} \\
\bottomrule
\end{tabular}
\tabnote{$^{\rm{a}}$ \cite{Walter+08}}
\tabnote{$^{\rm{b}}$ \cite{deBlok+08}}
\vspace{2mm}
\end{table}

To make a comparison, we use the natural-weighted \HI data cube of NGC\,3521 which is taken from THINGS \citep{Walter+08}. NGC\,3521 has a high inclination value of $i=72.7^{\circ}$ (\citealt{deBlok+08}). A galaxy with a high inclination is more affected by beam smearing. In addition, the projection effect of line-of-sight velocities is more severe for highly inclined galaxies. In this regard, NGC\,3521 is a suitable sample for which the two super-profile methods can be tested. The THINGS \HI data cube has dimensions of $1024 \times 1024$ pixels for {\sc ra} and {\sc dec} with a pixel scale of 1.5\arcsec, and 109 channels for the velocity axis. The \HI beam size and the channel resolution of the cube are $\sim$10\arcsec\ and 5.2~\kms, respectively.

\begin{figure}[!t]
\centering
\includegraphics[width=\linewidth]{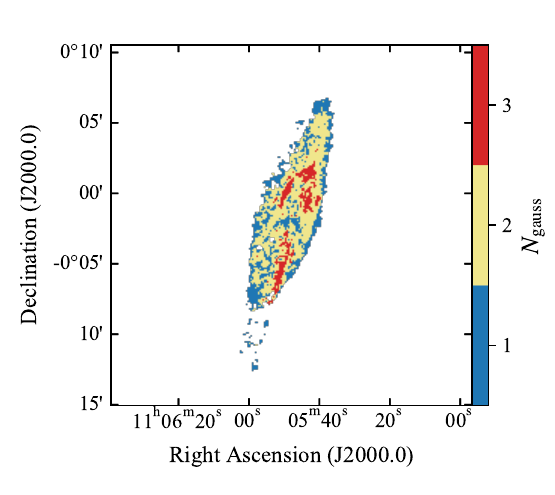}
\caption{The $N_{\rm{gauss}}$ map of NGC\,3521.\label{fig:ngauss}}
\vspace{2mm}
\end{figure}

\begin{figure*}[!t]
\centering
\includegraphics[width=1.0\textwidth]{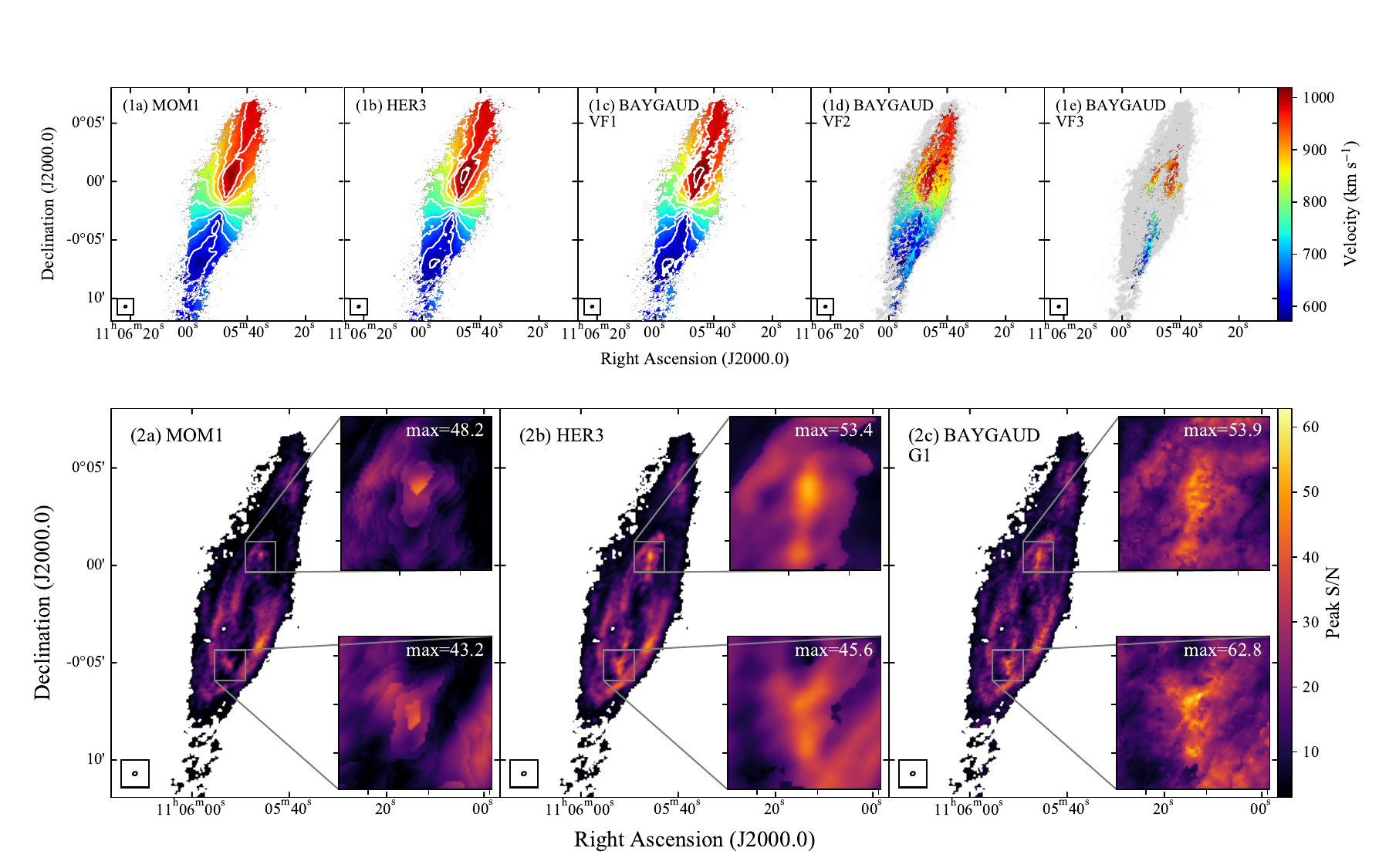}
\caption{The {\sc baygaud} profile analysis result for NGC\,3521. Upper panels compare centroid velocities estimated using different methods. Panel (1a): {\sc moment1} velocity field, (1b): Gauss-Hermite $h_3$ velocity field, and (1c) to (1e): {\sc baygaud} velocity fields. For pixels fitted with more than one Gaussian component, we sort each component's peak signal-to-noise ratio (S/N) into descending order and place it in respective order from {\sc vf1} to {\sc vf3}. The iso-velocity contour levels range from 572~\kms\ to 1,019~\kms\ in steps of 10~\kms. The bottom panels compare peak S/N calculated using (2a): {\sc moment1} analysis, (2b): Gauss-Hermite $h_{3}$ fit, and (2c): {\sc baygaud vf1}. The maximum values are shown in each respective zoomed inset. The circles on the bottom left side of each panel are the beam size of THINGS NGC\,3521.\label{fig:baygaud_peaksn}}
\vspace{2mm}
\end{figure*}

We run {\sc baygaud} to the data cube in order to decompose the \HI velocity profiles of NGC\,3521 with optimal numbers of Gaussian components. We let {\sc baygaud} fit a line profile with up to three Gaussian components. From a visual inspection of line profiles of the cube, three Gaussian components are found to be enough to model shapes of the most extreme non-Gaussian profiles. As described in Section~\ref{sec:s21}, {\sc baygaud} finds the most appropriate Gaussian model via the model selection criteria based on the Bayes factors for the tried models. We adopt the best model found only when all the S/N values of the decomposed Gaussian components for the line profile are greater than three. If not, we simply use the single Gaussian fitting result as the best model of the profile. In addition, we also derive the centroid velocities of the line profiles using the {\sc moment} analysis ({\sc moment1}) and a Hermite $h_3$ polynomial function as in \citetalias{Ianja+12}.

In Figure~\ref{fig:ngauss}, we show the $N_{\rm{gauss}}$ map of NGC\,3521 which is derived from the {\sc baygaud} profile decomposition. $N_{\rm{gauss}}$ is the number of optimally decomposed Gaussian components for each velocity profile. In most disk regions of NGC\,3521, the velocity profiles are best described by Gaussian models with $N_{\rm{gauss}} = 2$ while a single Gaussian model better describes the ones in the outer region. Interestingly, some of the profiles in the inner region which are located around the major axis are described by Gaussian models with $N_{\rm{gauss}} = 3$. This indicates that the gas kinematics in this region is complex and can be well decomposed with three components at different line-of-sight velocities. As discussed earlier, this could be due to observational systematic effects like beam smearing or projection effect or both. These effects are particularly enhanced in the inner \mbox{regions} along the kinematic major axis of a galaxy where the velocity gradient is the highest. Or this could be associated with hydrodynamical and/or gravitational processes in the galaxy like star formation and stellar feedback.

In Figure~\ref{fig:baygaud_peaksn}, we present the profile analysis results for the THINGS \HI data cube of NGC\,3521 derived using the three different profile analysis methods, {\sc baygaud}, Hermite $h_3$ polynomial fitting ({\sc her3}) and {\sc moment1} ({\sc mom1}). In the following, we describe the results shown in Figure~\ref{fig:baygaud_peaksn}:

\noindent {\bf 1) Upper panels---velocity fields:} 2D maps of the centroid velocities derived from the profile \mbox{analyses-1a}: {\sc mom1} ({\sc moment1}), 1b: {\sc her3} (Hermite $h_3$), 1c: {\sc baygaud-g1} (\mbox{{\sc baygaud}-1st} Gaussian component), 1d: {\sc baygaud-g2} (\mbox{{\sc baygaud}-2nd} Gaussian component), and 1e: {\sc baygaud-g3} ({{\sc baygaud}-3rd} Gaussian component). The iso-velocity contours range from 572 to 1,019~\kms\ in steps of 10~\kms. We run {\sc baygaud} with a maximum number of Gaussians of three, which provides the fitting results for a set of three Gaussian components. To make comparisons with the results from the {\sc moment} and Hermite $h_3$ analyses, we construct the {\sc baygaud} velocity field maps for the three Gaussian components (G1, G2 and G3) whose peak fluxes at a pixel position are sorted in descending order. These are denoted as {\sc vf1}, {\sc vf2} and {\sc vf3} in Figure~\ref{fig:baygaud_peaksn}.

\begin{table}
\centering
\caption{Mass fraction of NGC\,3521 of {\sc baygaud} components sorted in descending order of each component's peak S/N.\label{tab:mass_fraction}}

\begin{tabular}{y{0.5\linewidth} y{0.5\linewidth}}
\toprule
 & Mass fraction\\%
\midrule
{\sc baygaud} G1 & 0.679\\%
{\sc baygaud} G2 & 0.271\\%
{\sc baygaud} G3 & 0.050\\%
\bottomrule
\end{tabular}
\end{table}

\begin{figure*}[!t]
\centering
\includegraphics[width=\linewidth]{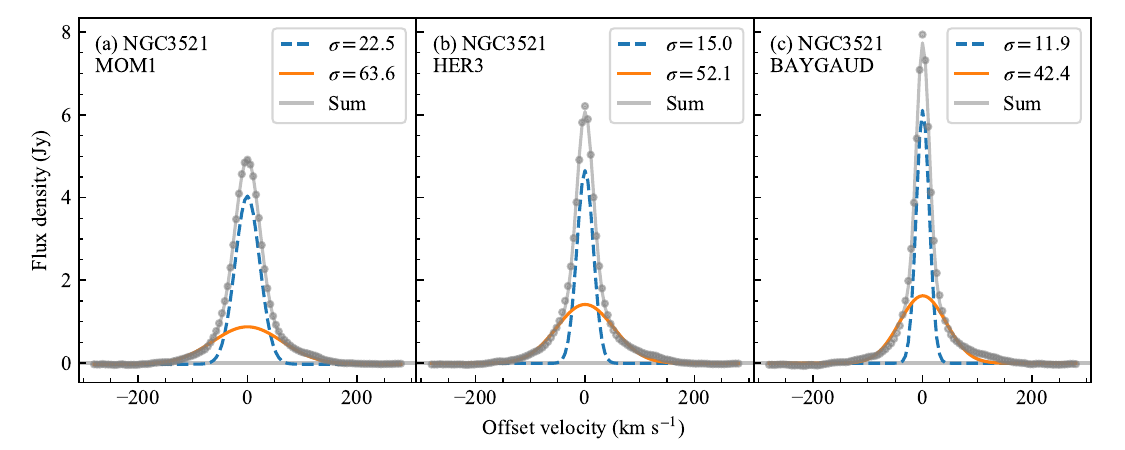}
\caption{Super-profiles of NGC\,3521 with {\sc 2gfit} parameterization results. Panel (a): The super-profile built using {\sc mom1} velocity, (b): the super-profile built using velocity from Gauss-Hermite $h_3$ fit, and (c): the super-profile built using our method. The grey dots with verticle lines show super-profile data points with their 3-$\sigma$ uncertainties although they are generally smaller than the markers. The blue line represents the narrow Gaussian component and the orange line represents the broad component. The gray line is the sum of narrow and broad Gaussian components. \label{fig:GFITs_momherbay}}
\vspace{2mm}
\end{figure*}

As shown in the upper panels of Figure~\ref{fig:ngauss}, the centroid velocities of the \HI velocity profiles of NGC\,3521 derived using the different methods are mostly consistent with each other except for the regions where multiple Gaussian components with $N_{\rm{gauss}}>1$ are present. The centroid velocities derived from the {\sc moment1} and hermite $h_3$ polynomial fitting analyses should be affected by these secondary or tertiary Gaussian components. The \HI mass fractions of the {\sc baygaud-G1}, {\sc -G2}, and {\sc -G3} components of the profiles are given in Table~\ref{tab:mass_fraction}. The \HI masses of the secondary and tertiary Gaussian components occupy $\sim$30\% of the total \HI mass.

\noindent {\bf 2) Lower panels\---S/N maps:} 2D maps of the peak flux S/N values of the velocity profiles of the cube derived from the profile \mbox{analyses-2a}: {\sc MOM1} ({\sc moment} analysis), 2b: {\sc HER3} (Hermite $h_3$ polynomial fitting), 2c: {\sc baygaud-g1} ({\sc baygaud} analysis). For the case of the {\sc baygaud} analysis, we only show the peak S/N values for the {\sc baygaud}-G1 component whose peak fluxes are higher than those of the {\sc baygaud}-G2, and -G3 ones. The peak flux values for the {\sc baygaud} and Hermite $h_3$ polynomial fitting analyses are computed using their best-fit parameters while the ones for the {\sc moment} analysis are directly taken from the observed flux values of line profiles at the corresponding centroid velocities measured from the {\sc moment1}. For the noise levels of the profiles, we use their rms noises in the line-free channels of the cube.

As shown in the zoomed-in inset panels of the figure, the peak fluxes derived from the {\sc baygaud} analysis are higher than those from both the {\sc moment} and Hermite $h_3$ polynomial fitting methods. On the other hand, the ones from the {\sc moment} analysis are the lowest. This is caused by asymmetric shapes of the line profiles, which is particularly prominent in the regions with $N_{\rm{gauss}}>1$ as shown in the inset panels. As discussed earlier, the intensity-weighted mean velocity for a line profile deviates from its peak flux velocity if additional kinematic components are present in the profile. The Hermite $h_3$ fitting method also has limitations in deriving the centroid velocity close to the peak flux velocity of a profile unless the profile shape is symmetric. The velocity deviation becomes significant as the amplitudes of the additional components increase. This results in uncertainties of the peak flux values of line profiles. However, such a bias can be largely reduced in the {\sc baygaud} analysis which explicitly models the additional kinematic components using a set of Gaussian functions.

\subsection{H\textsc{i} Super-Profiles of NGC 3521\label{sec:s32}}

We then fit a double Gaussian model to the super-profiles to quantify their shapes. From this {\sc 2gfit} parameterization, we decompose the super-profiles with the narrower (smaller velocity dispersion) and broader (larger velocity dispersion) Gaussian components as presented in Figure~\ref{fig:GFITs_momherbay}. As shown in the figure, the {\sc baygaud}-based \HI super-profile has a narrower wing and higher peak in shape than the others. The integrated intensity (i.e., the area) of the narrower Gaussian component of the {\sc baygaud}-based super-profile is evidently higher than those of the others. Accordingly, the fraction of the integrated intensity of the broader Gaussian component is smaller than those of the others. The velocity dispersions of the narrower and broader Gaussian components of the {\sc baygaud}-based super-profile are smaller than the others.

According to star formation models (e.g., \citealt{Krumholz12}), atomic hydrogen gas should have cooled and thus passed a kinematically cool phase with a lower velocity dispersion before turning into molecular hydrogen gas, H$_{2}$. The kinematically narrower Gaussian component of an \HI super-profile can be associated with \HI gas in such a cool phase. In this regard, the {\sc baygaud}-based \HI super-profile is superior to extract more kinematically cool \HI gas in galaxies.

\begin{figure*}[!t]
\centering
\includegraphics[width=\linewidth]{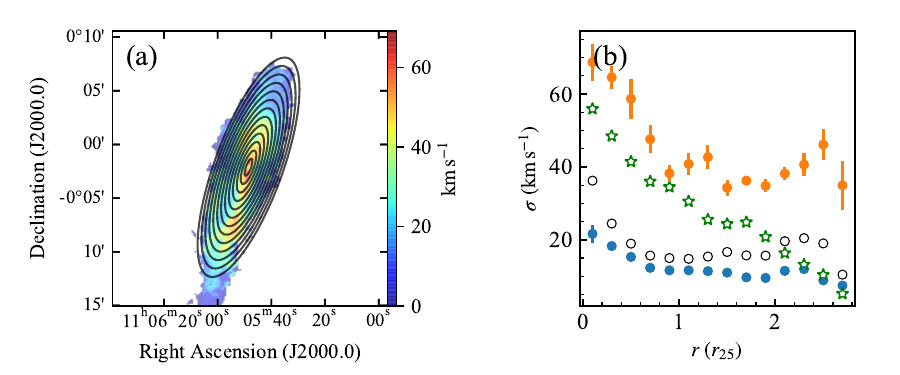}
\caption{The ring-by-ring variation of velocity dispersion of NGC\,3521. Panel (a): {\sc moment2} map of NGC\,3521 overplotted with 0.2 $r_{25}$ width rings, and panel (b): ring-by-ring variation of velocity dispersions; blue and orange solid circles: narrow and broad components from {\sc 2gfit} on each super-profiles, black open circles: velocity dispersion of single Gaussian fit on each super-profiles, and green stars: median of {\sc moment2} values in each ring.}\label{fig:n3521_rings}
\vspace{2mm}
\end{figure*}

The stacking method can also be used to study the radial variation of velocity dispersion of a galaxy by stacking in concentric rings (e.g., \citealt{Ianja+15, Das+20}). 
In Figure~\ref{fig:n3521_rings}, we show the radial variation of velocity dispersion of NGC\,3521 derived by stacking each concentric rings defined by adopting constant position angle and inclination from \cite{deBlok+08}. The ring widths of $0.2 \ r_{25}$ is adopted, following \cite{Ianja+15}. Panel (a) shows the {\sc moment2} map of NGC\,3521 overplotted with rings and panel (b) shows the ring-by-ring variation of velocity dispersions. The circles show the velocity dispersions from the stacking method; the blue and orange sold circles show the velocity dispersions of the narrow and the broad components from the double Gaussian model ({\sc 2gfit}), and black open circles show the velocity dispersion derived by fitting single Gaussian model. The green star symbols show the median of {\sc moment2} values in each ring. We see a similar decline of velocity dispersions traced with {\sc 2gfit} with increasing radius as in \cite{Ianja+15}.

\section{Practical Application of the Baygaud-Based Hi Super-Profile to Nearby Galaxies\label{sec:s4}}

In this Section, we make a practical application of the {\sc baygaud}-based \HI super-profile analysis to nearby galaxies in the local Universe ($\lesssim$15~Mpc) taken from THINGS and LITTLE THINGS. We then correlate the \HI super-profiles’ shape parameters with physical properties of the galaxies like SFR and discuss how comparable the correlations derived from the {\sc baygaud}-based \HI super-profiles are to those from the other two; {\sc sp} and {\sc her3}-based super-profiles.

\begin{table*}[!hp]
\caption{Basic properties of sample galaxies.\label{tab:info_samples}}
\centering\small
\begin{tabular}{x{0.125\linewidth} y{0.2\linewidth} *{2}{y{0.137\linewidth}} *{4}{y{0.1\linewidth}}}
\toprule
\quad Galaxy & Survey & $\alpha$ (2000.0) & $\delta$ (2000.0) & $D$ & $i_{\rm{HI}}$ & log $M_{\rm{HI}}$ & log SFR\\
& & [$^{\rm{h}}$ $^{\rm{m}}$ $^{\rm{s}}$]  & [$^{\circ}$ $'$ \arcsec]  & [Mpc] & [$^{\circ}$] & [$M_{\odot}$] & [$M_{\odot}$\,$\rm{yr}^{-1}$]\\
& (1) & (2) & (3) & (4) & (5) & (6) & (7)\\
\midrule
\quad CVnIdwA & LITTLE THINGS & 12 38 40.1 & $+$32 45 59.0 & \pt{0}3.6 & 66.5 & \pt{0}7.67 & $-2.47$\\
\quad DDO\,43 & LITTLE THINGS & 07 28 17.7 & $+$40 46 11.4 & \pt{0}7.8 & 40.6 & \pt{0}8.23 & $-1.83$\\
\quad DDO\,46 & LITTLE THINGS & 07 41 25.7 & $+$40 06 46.1 & \pt{0}6.1 & 27.9 & \pt{0}8.27 & $-1.85$\\
\quad DDO\,47 & LITTLE THINGS & 07 41 56.3 & $+$16 48 07.4 & \pt{0}5.2 & 45.5 & \pt{0}8.59 & $-1.63$\\
\quad DDO\,50$^{\dagger}$ & LITTLE THINGS & 08 19 04.9 & $+$70 43 13.1 & \pt{0}3.4 & 49.7 & \pt{0}8.85 & $-0.97$\\ 
\quad DDO\,52 & LITTLE THINGS & 08 28 28.6 & $+$41 51 27.1 & 10.3 & 43.0 & \pt{0}8.43 & $-1.70$ \\ 
\quad DDO\,53$^{\dagger}$ & LITTLE THINGS & 08 34 07.3 & $+$66 10 54.6 & \pt{0}3.6 & 27.0 & \pt{0}7.72 & $-2.20$ \\ 
\quad DDO\,63$^{\dagger}$ & LITTLE THINGS & 09 40 32.3 & $+$71 10 56.0 & \pt{0}3.9 & 13.0\rlap{$^{\rm{a}}$} & \pt{0}8.19 & $-1.79$ \\ 
\quad DDO\,69 & LITTLE THINGS & 09 59 26.5 & $+$30 44 47.0 & \pt{0}0.8 & ... & \pt{0}6.84 & $-3.17$ \\ 
\quad DDO\,70 & LITTLE THINGS & 10 00 00.8 & $+$05 20 09.5 & \pt{0}1.3 & 50.0 & \pt{0}7.61 & $-2.30$ \\ 
\quad DDO\,75 & LITTLE THINGS & 10 11 00.5 & $-$04 41 30.0 & \pt{0}1.3 & ... & \pt{0}7.86 & $-1.89$ \\ 
\quad DDO\,87 & LITTLE THINGS & 10 49 36.4 & $+$65 32 01.1 & \pt{0}7.7 & 55.5 & \pt{0}8.39 & $-0.27$ \\ 
\quad DDO\,101 & LITTLE THINGS & 11 55 39.7 & $+$31 31 06.5 & \pt{0}6.4 & 51.0 & \pt{0}7.36 & $-2.37$ \\ 
\quad DDO\,126 & LITTLE THINGS & 12 27 05.6 & $+$37 08 30.4 & \pt{0}4.9 & 65.0 & \pt{0}8.16 & $-1.72$ \\ 
\quad DDO\,133 & LITTLE THINGS & 12 32 54.5 & $+$31 32 30.0 & \pt{0}3.5 & 43.4 & \pt{0}8.02 & $-1.94$ \\ 
\quad DDO\,154$^{\dagger}$ & LITTLE THINGS & 12 54 05.2 & $+$27 08 58.7 & \pt{0}3.7 & 68.2 & \pt{0}8.46 & $-1.89$\\
\quad DDO\,155 & LITTLE THINGS & 12 58 40.2 & $+$14 13 00.1 & \pt{0}2.2 & ... & \pt{0}7.00 & $-2.59$\rlap{*}\\
\quad DDO\,165 & LITTLE THINGS & 13 06 24.9 & $+$67 42 25.0 & \pt{0}4.6 & ... & \pt{0}8.13 & $-2.46$\rlap{*}\\
\quad DDO\,167 & LITTLE THINGS & 13 13 22.8 & $+$46 19 21.7 & \pt{0}4.2 & ... & \pt{0}7.17 & $-2.30$\\
\quad DDO\,168 & LITTLE THINGS & 13 14 28.0 & $+$45 55 11.1 & \pt{0}4.3 & 46.5 & \pt{0}8.47 & $-1.72$\\
\quad DDO\,187 & LITTLE THINGS & 14 15 56.7 & $+$23 03 16.2 & \pt{0}2.2 & ... & \pt{0}7.12 & $-2.97$\\
\quad DDO\,210 & LITTLE THINGS & 20 46 51.7 & $-$12 50 54.0 & \pt{0}0.9 & 66.7 & \pt{0}6.30 & $-3.75$\\
\quad DDO\,216 & LITTLE THINGS & 23 28 35.2 & $+$14 44 35.0 & \pt{0}1.1 & 63.7 & \pt{0}6.75 & $-3.25$\\
\quad F564-V3 & LITTLE THINGS & 09 02 53.4 & $+$20 04 32.2 & \pt{0}8.7 & 56.5 & \pt{0}7.61 & $-2.84$\\
\quad Haro\,29 & LITTLE THINGS & 12 26 16.2 & $+$48 29 36.6 & \pt{0}5.8 & 61.2 & \pt{0}7.80 & $-1.65$\\
\quad Haro\,36 & LITTLE THINGS & 12 46 56.4 & $+$51 36 46.8 & \pt{0}9.3 & 70.0 & \pt{0}8.16 & $-1.38$\\
\quad IC\,10 & LITTLE THINGS & 00 20 23.2 & $+$59 17 34.7 & \pt{0}0.7 & 47.0 & \pt{0}7.78 & $-1.41$\rlap{*}\\
\quad IC\,1613 & LITTLE THINGS & 01 04 54.2 & $+$02 08 00.0 & \pt{0}0.7 & 48.0 & \pt{0}7.53 & $-1.97$\\ 
\quad LGS\,3 & LITTLE THINGS & 01 03 56.0 & $+$21 53 41.0 & \pt{0}0.7 & ... & \pt{0}5.19 & $-4.66$\\
\quad M81\,dwA$^{\dagger}$ & LITTLE THINGS & 08 23 55.1 & $+$71 01 56.0 & \pt{0}3.6 & ... & \pt{0}7.18 & $-2.93$\\
\quad M81\,dwB$^{\dagger}$ & THINGS & 10 05 30.6 & $+$70 21 52.0 & \pt{0}5.3 & 44.0\rlap{$^{\rm{a}}$} & \pt{0}7.40 & $-2.30$\\
\quad Mrk\,178 & LITTLE THINGS & 11 33 29.1 & $+$49 14 17.4 & \pt{0}3.9 & ... & \pt{0}6.99 & $-2.13$\\
\quad NGC\,628$^{\dagger}$ & THINGS & 01 36 41.8 & $+$15 47 00.0 & \pt{0}7.3 & ... & \pt{0}9.58 & \pt{+}$0.08$\\
\quad NGC\,925$^{\dagger}$ & THINGS & 02 27 16.5 & $+$33 34 44.0 & \pt{0}9.2 & 63.8 & \pt{0}9.66 & \pt{+}$0.04$\\
\quad NGC\,1569$^{\dagger}$ & LITTLE THINGS & 04 30 49.2 & $+$64 50 52.5 & \pt{0}3.4 & 69.1 & \pt{0}8.39 & $-0.35$\\
\quad NGC\,2366$^{\dagger}$ & LITTLE THINGS & 07 28 51.2 & $+$69 12 31.1 & \pt{0}3.4 & 63.0 & \pt{0}8.84 & $-0.90$\\
\quad NGC\,2403 & THINGS & 07 36 51.1 & $+$65 36 03.0 & \pt{0}3.2 & 62.9 & \pt{0}9.41 & $-0.07$\\
\quad NGC\,2841$^{\dagger}$ & THINGS & 09 22 02.6 & $+$50 58 35.0 & 14.1 & 73.7 & \pt{0}9.93 & $-0.70$\\
\quad NGC\,2903$^{\dagger}$ & THINGS & 09 32 10.1 & $+$21 30 04.0 & \pt{0}8.9 & 65.2 & \pt{0}9.64 & $-0.40$\rlap{$^{\rm{b}}$}\\
\quad NGC\,2976$^{\dagger}$ & THINGS & 09 47 15.3 & $+$67 55 00.0 & \pt{0}3.6 & 64.5 & \pt{0}8.13 & $-1.00$\\
\quad NGC\,3031 & THINGS & 09 55 33.1 & $+$69 03 55.0 & \pt{0}3.6 & 59.0 & \pt{0}9.56 & $0.03$\\
\quad NGC\,3077$^{\dagger}$ & THINGS & 10 03 19.1 & $+$68 44 02.0 & \pt{0}3.8 & ... & \pt{0}8.94 & $-1.05$\\
\quad NGC\,3184$^{\dagger}$ & THINGS & 10 18 17.0 & $+$41 25 28.0 & 11.1 & ... & \pt{0}9.49 & \pt{+}$0.16$\\
\quad NGC\,3198$^{\dagger}$ & THINGS & 10 19 55.0 & $+$45 32 59.0 & 13.8 & 71.5 & 10.01 & $-0.07$\\
\quad NGC\,3351$^{\dagger}$ & THINGS & 10 43 57.7 & $+$11 42 14.0 & 10.1 & ... & \pt{0}9.08 & $-0.15$\\
\quad NGC\,3521$^{\dagger}$ & THINGS & 11 05 48.6 & $-$00 02 09.0 & 10.7 & 72.7 & \pt{0}9.90 & \pt{+}$0.52$\\
\quad NGC\,3621$^{\dagger}$ & THINGS & 11 18 16.5 & $-$32 48 51.0 & \pt{0}6.6 & 64.7 & \pt{0}9.85 & \pt{+}$0.32$\\
\quad NGC\,3627$^{\dagger}$ & THINGS & 11 20 15.0 & $+$12 59 30.0 & \pt{0}9.3 & 61.8 & \pt{0}8.91 & \pt{+}$0.39$\\
\bottomrule
\end{tabular}
\tabnote{
(1): Name of the survey; (2), (3): Position, taken from {\sc simbad}$^{\rm{d}}$ for LITTLE THINGS; (4): Distance; (5): \HI inclination from 2D tilted ring analysis, taken from \cite{deBlok+08} for THINGS and \cite{Oh+15} for LITTLE THINGS; (6): \HI mass; (7): Global star formation rate, get from H$\alpha$ for THINGS and FUV for LITTLE THINGS unless marked with *, which we used SFR$_{\rm{H}\alpha}$ instead when SFR$_{\rm{FUV}}$ information was not available; Unless mentioned otherwise, values are taken from \cite{Walter+08} for THINGS and \cite{Hunter+12} for LITTLE THINGS. Galaxies marked with $\dagger$ are also used in the super-profile analysis by \citetalias{Ianja+12}. ``..." indicates that no data is available. \\
$^{\rm{a}}$ \cite{Oh+11}; $^{\rm{b}}$ \cite{Popping+10}; $^{\rm{c}}$ \cite{Hunter+99}, \cite{Cignoni+18}; $^{\rm{d}}$ \url{http://simbad.u-strasbg.fr/simbad/}, \cite{Wenger+00}.
}
\end{table*}

\begin{table*}[!t]
\ContinuedFloat
\caption{Continued}
\centering\small
\begin{tabular}{x{0.125\linewidth} y{0.2\linewidth} *{2}{y{0.137\linewidth}} *{4}{y{0.1\linewidth}}}
\toprule
\quad Galaxy & Survey & $\alpha$ (2000.0) & $\delta$ (2000.0) & $D$ & $i_{\rm{HI}}$ & log $M_{\rm{HI}}$ & log SFR\\
& & [$^{\rm{h}}$ $^{\rm{m}}$ $^{\rm{s}}$]  & [$^{\circ}$ $'$ \arcsec]  & [Mpc] & [$^{\circ}$] & [$M_{\odot}$] & [$M_{\odot}$\,$\rm{yr}^{-1}$]\\
& (1) & (2) & (3) & (4) & (5) & (6) & (7)\\
\midrule
\quad NGC\,3738 & LITTLE THINGS & 11 35 49.0 & $+$54 31 24.7 & \pt{0}4.9 & 22.6 & 8.06 & $-1.25$\\
\quad NGC\,4163 & LITTLE THINGS & 12 12 09.1 & $+$36 10 02.8 & \pt{0}2.9 & ... & 7.16 & $-2.38$\\
\quad NGC\,4214$^{\dagger}$ & LITTLE THINGS & 12 15 39.2 & $+$36 19 36.8 & \pt{0}3.0 & ... & 8.76 & $-0.83$\\
\quad NGC\,4449$^{\dagger}$ & THINGS & 12 28 11.9 & $+$44 05 40.0 & \pt{0}4.2 & ... & 9.04 & $-0.31$\rlap{$^{\rm{c}}$}\\
\quad NGC\,4736$^{\dagger}$ & THINGS & 12 50 53.0 & $+$41 07 13.0 & \pt{0}4.7 & 41.4 & 8.60 & $-0.37$\\
\quad NGC\,4826$^{\dagger}$ & THINGS & 12 56 43.6 & $+$21 41 00.0 & \pt{0}7.5 & 65.2 & 8.74 & $-0.09$\\
\quad NGC\,5055$^{\dagger}$ & THINGS & 13 15 49.2 & $+$42 01 45.0 & 10.1 & 59.0 & 9.96 & \pt{+}$0.38$\\
\quad NGC\,5194$^{\dagger}$ & THINGS & 13 29 52.7 & $+$47 11 43.0 & \pt{0}8.0 & ... & 9.40 & \pt{+}$0.78$\\
\quad NGC\,5236$^{\dagger}$ & THINGS & 13 37 00.9 & $-$29 51 57.0 & \pt{0}4.5 & ... & 9.23 & \pt{+}$0.40$\\
\quad NGC\,5457$^{\dagger}$ & THINGS & 14 03 12.6 & $+$54 20 57.0 & \pt{0}7.4 & ... & 10.15 & \pt{+}$0.40$\\
\quad NGC\,6946$^{\dagger}$ & THINGS & 20 34 52.2 & $+$60 09 14.0 & \pt{0}5.9 & 32.6 & 9.62 & \pt{+}$0.68$\\
\quad NGC\,7331$^{\dagger}$ & THINGS & 22 37 04.1 & $+$34 24 57.0 & 14.7 & 75.8 & 9.96 & \pt{+}$0.62$\\
\quad NGC\,7793$^{\dagger}$ & THINGS & 23 57 49.7 & $-$32 35 28.0 & \pt{0}3.9 & 49.6 & 8.95 & $-0.29$\\
\quad SagDIG & LITTLE THINGS & 19 29 59.0 & $-$17 40 41.0 & \pt{0}1.1 & ... & 6.94 & $-2.89$\\
\quad UGC\,8508 & LITTLE THINGS & 13 30 44.9 & $+$54 54 38.5 & \pt{0}2.6 & 82.5 & 7.28 & $-2.67$\rlap{*}\\
\quad VIIZw403 & LITTLE THINGS & 11 28 00.4 & $+$78 59 38.4 & \pt{0}4.4 & ... & 7.69 & $-1.74$\\
\quad WLM & LITTLE THINGS & 00 01 57.9 & $-$15 27 50.0 & \pt{0}1.0 & 74.0 & 7.85 & $-2.04$\\
\bottomrule
\end{tabular}
\vspace{2mm}
\end{table*}

\subsection{Sample Galaxies and H\textsc{i} Data\label{sec:s41}}

We select 64 sample galaxies from THINGS and LITTLE THINGS whose star formation rate values are available. The observational and physical properties of the sample galaxies are presented in Table~\ref{tab:info_samples}. Of these, 31 sample galaxies, marked with a $\dagger$ symbol, were also used for the \HI super-profile analysis in \citeauthor{Ianja+12} (\citeyear{Ianja+12}; see also \citeyear{Ianja+15}).
To construct \HI super-profiles of the sample galaxies we use the natural weighted \HI data cubes of nearby galaxies taken from THINGS (\citealt{Walter+08}) and LITTLE THINGS (\citealt{Hunter+12}). Both surveys provide high spatial ($\sim$10\arcsec) and spectral ($<$5.2~\kms) resolution \HI data cubes of the galaxies. The corresponding physical resolutions range from $\sim$30~pc to $\sim$870~pc with a mean of $\sim$300~pc which is high enough to resolve individual giant molecular gas clouds (GMCs) in the galaxies. Moreover, the observations with an average of 12~hours integration time for each galaxy provide high-quality \HI data cubes with good S/N throughout the gas disk of the galaxies. These high-quality \HI data cubes are useful for testing our \HI super-profile method.

\subsection{H\textsc{i} Super-Profiles of the Sample Galaxies\label{sec:s42}}

\begin{table*}[!p]
\caption{{\sc 2gfit} parameterization results for super-profile sample.\label{tab:GFIT_super-profiles}}
\centering\small
\begin{tabular}{x{0.166\linewidth} *{5}{y{0.166\linewidth}}}
\toprule
\qquad Galaxy & $\sigma_{\rm{n}}$ & $\sigma_{\rm{b}}$ & \snsb & \anab & \anat\\
       & [\kms] & [\kms]\\
       & (1) & (2) & (3) & (4) & (5)\\
\midrule
\multicolumn{6}{c}{Clean sample (47 in total)}\\
\midrule
\qquad CVnIdwA & \pt{0}$5.1\pm0.6$ & \pt{0}$9.5\pm1.4$ & $0.53\pm0.10$ & $0.08\pm0.03$ & $0.08\pm0.03$ \\ 
\qquad DDO\,43 & \pt{0}$5.6\pm0.6$ & \pt{0}$9.9\pm0.4$ & $0.56\pm0.06$ & $0.31\pm0.10$ & $0.24\pm0.07$ \\ 
\qquad DDO\,46 & \pt{0}$6.1\pm0.7$ & $10.9\pm1.7$ & $0.56\pm0.10$ & $0.24\pm0.09$ & $0.19\pm0.07$ \\ 
\qquad DDO\,47 & \pt{0}$7.0\pm0.2$ & $14.1\pm0.7$ & $0.49\pm0.03$ & $1.05\pm0.15$ & $0.51\pm0.05$ \\ 
\qquad DDO\,50 & \pt{0}$6.1\pm0.2$ & $13.7\pm0.5$ & $0.45\pm0.02$ & $0.63\pm0.06$ & $0.39\pm0.03$ \\ 
\qquad DDO\,53 & \pt{0}$7.2\pm0.3$ & $14.2\pm0.8$ & $0.51\pm0.04$ & $0.95\pm0.19$ & $0.49\pm0.07$ \\ 
\qquad DDO\,63 & \pt{0}$6.6\pm0.4$ & $14.3\pm4.6$ & $0.46\pm0.15$ & $0.93\pm0.35$ & $0.48\pm0.10$ \\ 
\qquad DDO\,69 & \pt{0}$3.9\pm0.3$ & \pt{0}$8.8\pm0.5$ & $0.45\pm0.04$ & $0.41\pm0.06$ & $0.29\pm0.04$ \\ 
\qquad DDO\,70 & \pt{0}$5.1\pm0.2$ & $11.8\pm0.3$ & $0.43\pm0.02$ & $0.33\pm0.02$ & $0.25\pm0.02$ \\ 
\qquad DDO\,75 & \pt{0}$5.4\pm0.1$ & $11.9\pm0.2$ & $0.45\pm0.01$ & $0.36\pm0.02$ & $0.26\pm0.02$ \\ 
\qquad DDO\,126 & \pt{0}$5.9\pm0.5$ & $11.0\pm6.3$ & $0.54\pm0.31$ & $0.35\pm0.22$ & $0.26\pm0.13$ \\ 
\qquad DDO\,133 & \pt{0}$6.7\pm0.2$ & $13.6\pm0.6$ & $0.49\pm0.03$ & $0.98\pm0.13$ & $0.50\pm0.05$ \\ 
\qquad DDO\,154 & \pt{0}$7.5\pm0.2$ & $13.7\pm0.7$ & $0.54\pm0.03$ & $1.24\pm0.26$ & $0.55\pm0.07$ \\ 
\qquad DDO\,155 & \pt{0}$6.4\pm0.3$ & $14.4\pm0.8$ & $0.45\pm0.03$ & $0.74\pm0.12$ & $0.43\pm0.05$ \\ 
\qquad DDO\,167 & \pt{0}$5.0\pm0.4$ & $12.7\pm0.6$ & $0.39\pm0.04$ & $0.39\pm0.06$ & $0.28\pm0.04$ \\ 
\qquad DDO\,187 & \pt{0}$6.1\pm0.4$ & $13.6\pm0.4$ & $0.45\pm0.03$ & $0.32\pm0.05$ & $0.24\pm0.03$ \\ 
\qquad DDO\,210 & \pt{0}$5.0\pm0.2$ & $10.3\pm0.4$ & $0.49\pm0.03$ & $0.78\pm0.11$ & $0.44\pm0.05$ \\ 
\qquad DDO\,216 & \pt{0}$4.0\pm0.2$ & \pt{0}$8.9\pm0.2$ & $0.45\pm0.03$ & $0.47\pm0.05$ & $0.32\pm0.03$ \\ 
\qquad F564-V3 & \pt{0}$7.0\pm0.6$ & $12.7\pm0.4$ & $0.55\pm0.05$ & $0.27\pm0.07$ & $0.21\pm0.05$ \\ 
\qquad Haro\,29 & \pt{0}$5.9\pm0.8$ & $11.8\pm0.4$ & $0.50\pm0.07$ & $0.18\pm0.06$ & $0.16\pm0.05$ \\ 
\qquad Haro\,36 & $10.8\pm0.6$ & $23.9\pm0.9$ & $0.45\pm0.03$ & $0.42\pm0.06$ & $0.29\pm0.04$ \\ 
\qquad IC\,10 & \pt{0}$5.3\pm0.3$ & $12.1\pm0.4$ & $0.44\pm0.03$ & $0.55\pm0.06$ & $0.36\pm0.03$ \\ 
\qquad IC\,1613 & \pt{0}$3.7\pm0.3$ & \pt{0}$8.2\pm6.0$ & $0.45\pm0.33$ & $0.59\pm0.46$ & $0.37\pm0.19$ \\ 
\qquad M81dwA & \pt{0}$5.2\pm0.5$ & $12.9\pm0.8$ & $0.40\pm0.05$ & $0.39\pm0.08$ & $0.28\pm0.05$ \\ 
\qquad M81dwB & \pt{0}$5.7\pm0.3$ & $15.0\pm0.4$ & $0.38\pm0.02$ & $0.20\pm0.02$ & $0.16\pm0.01$ \\ 
\qquad Mrk\,178 & \pt{0}$5.9\pm0.2$ & $16.4\pm0.5$ & $0.36\pm0.02$ & $0.29\pm0.02$ & $0.23\pm0.02$ \\ 
\qquad NGC\,925 & \pt{0}$9.2\pm0.4$ & $24.0\pm1.8$ & $0.39\pm0.03$ & $0.93\pm0.19$ & $0.48\pm0.06$ \\ 
\qquad NGC\,2366 & \pt{0}$8.1\pm0.2$ & $18.7\pm0.5$ & $0.43\pm0.01$ & $0.87\pm0.09$ & $0.47\pm0.03$ \\ 
\qquad NGC\,2841 & $10.5\pm0.2$ & $40.7\pm1.8$ & $0.26\pm0.01$ & $0.78\pm0.08$ & $0.44\pm0.03$ \\ 
\qquad NGC\,2903 & \pt{0}$9.1\pm0.3$ & $29.5\pm1.6$ & $0.31\pm0.02$ & $0.90\pm0.12$ & $0.47\pm0.04$ \\ 
\qquad NGC\,2976 & \pt{0}$8.8\pm0.3$ & $21.2\pm3.3$ & $0.41\pm0.07$ & $1.02\pm0.20$ & $0.50\pm0.05$ \\ 
\qquad NGC\,3184 & \pt{0}$6.6\pm0.1$ & $18.6\pm0.3$ & $0.35\pm0.01$ & $0.45\pm0.02$ & $0.31\pm0.01$ \\ 
\qquad NGC\,3198 & \pt{0}$9.0\pm0.3$ & $21.9\pm0.7$ & $0.41\pm0.02$ & $0.85\pm0.10$ & $0.46\pm0.04$ \\ 
\qquad NGC\,3351 & \pt{0}$7.8\pm0.3$ & $22.6\pm3.5$ & $0.34\pm0.05$ & $1.09\pm0.21$ & $0.52\pm0.06$ \\ 
\qquad NGC\,3521 & $11.9\pm0.3$ & $42.4\pm2.8$ & $0.28\pm0.02$ & $1.05\pm0.13$ & $0.51\pm0.04$ \\ 
\qquad NGC\,3621 & \pt{0}$8.4\pm0.3$ & $22.9\pm1.3$ & $0.37\pm0.02$ & $1.00\pm0.16$ & $0.50\pm0.05$ \\ 
\qquad NGC\,3627 & $14.3\pm0.6$ & $43.5\pm3.1$ & $0.33\pm0.03$ & $0.84\pm0.13$ & $0.46\pm0.05$ \\ 
\qquad NGC\,4163 & \pt{0}$5.4\pm0.3$ & $11.3\pm0.3$ & $0.48\pm0.03$ & $0.21\pm0.03$ & $0.17\pm0.02$ \\ 
\qquad NGC\,4214 & \pt{0}$5.3\pm0.1$ & $13.0\pm0.2$ & $0.41\pm0.01$ & $0.48\pm0.02$ & $0.32\pm0.01$ \\ 
\qquad NGC\,4736 & \pt{0}$7.9\pm0.2$ & $22.2\pm1.0$ & $0.36\pm0.02$ & $1.01\pm0.10$ & $0.50\pm0.03$ \\ 
\qquad NGC\,4826 & \pt{0}$8.7\pm0.1$ & $29.1\pm0.8$ & $0.30\pm0.01$ & $1.00\pm0.05$ & $0.50\pm0.02$ \\ 
\qquad NGC\,5194 & $10.3\pm0.2$ & $28.3\pm0.9$ & $0.36\pm0.01$ & $0.55\pm0.03$ & $0.35\pm0.02$ \\ 
\qquad NGC\,7331 & $11.7\pm0.3$ & $31.5\pm0.7$ & $0.37\pm0.01$ & $0.64\pm0.04$ & $0.39\pm0.02$ \\ 
\qquad NGC\,7793 & \pt{0}$6.9\pm0.1$ & $17.6\pm0.3$ & $0.39\pm0.01$ & $0.67\pm0.04$ & $0.40\pm0.02$ \\ 
\qquad SagDIG & \pt{0}$5.1\pm0.1$ & $10.7\pm0.2$ & $0.47\pm0.02$ & $0.49\pm0.04$ & $0.33\pm0.02$ \\ 
\qquad VIIZw 403 & \pt{0}$8.6\pm0.7$ & $16.5\pm0.8$ & $0.52\pm0.05$ & $0.44\pm0.11$ & $0.30\pm0.07$ \\ 
\qquad WLM & \pt{0}$5.2\pm0.2$ & $11.4\pm0.4$ & $0.45\pm0.03$ & $0.75\pm0.10$ & $0.43\pm0.04$ \\ 
\bottomrule
\end{tabular}
\tabnote{(1): Velocity dispersion of narrow Gaussian component; (2): Velocity dispersion of broad Gaussian component; (3): Ratio of velocity dispersion of the narrow component and that of the broad component; (4): Ratio of area of narrow and broad Gaussian component; (5): Ratio of area of narrow Gaussian component and total Gaussian area (\at = \an + \ab).}
\end{table*}

\begin{table*}[!t]
\ContinuedFloat
\caption{Continued}
\centering\small
\begin{tabular}{x{0.166\linewidth} *{5}{y{0.166\linewidth}}}
\toprule
\qquad Galaxy & $\sigma_{\rm{n}}$ & $\sigma_{\rm{b}}$ & \snsb & \anab & \anat\\
       & [\kms] & [\kms]\\
       & (1) & (2) & (3) & (4) & (5)\\
\midrule
\multicolumn{6}{c}{Super-profile with a negative bowl or defects (17 in total)}\\
\midrule
\qquad DDO\,52 & \pt{0}$2.8\pm2.2$ & \pt{0}$8.1\pm4.7$ & $0.34\pm0.37$ & $0.03\pm0.14$ & $0.03\pm0.13$\\
\qquad DDO\,87 & \pt{0}$3.4\pm1.1$ & \pt{0}$7.7\pm2.5$ & $0.44\pm0.21$ & $0.05\pm0.09$ & $0.04\pm0.09$\\
\qquad DDO\,101 & \pt{0}$8.3\pm0.2$ & $37.6\pm3.5$ & $0.22\pm0.02$ & $1.62\pm0.26$ & $0.62\pm0.04$\\
\qquad DDO\,165 & \pt{0}$9.4\pm0.9$ & $23.6\pm3.2$ & $0.40\pm0.07$ & $1.04\pm0.41$ & $0.51\pm0.13$\\
\qquad DDO\,168 & \pt{0}$4.5\pm1.2$ & $10.1\pm1.5$ & $0.44\pm0.13$ & $0.06\pm0.07$ & $0.06\pm0.07$\\
\qquad LGS\,3 & \pt{0}$4.1\pm0.5$ & $10.7\pm6.0$ & $0.38\pm0.22$ & $0.47\pm0.28$ & $0.32\pm0.14$\\
\qquad NGC\,628 & \pt{0}$4.0\pm0.6$ & \pt{0}$9.7\pm0.4$ & $0.41\pm0.07$ & $0.13\pm0.04$ & $0.11\pm0.04$\\
\qquad NGC\,1569 & \pt{0}$7.4\pm1.3$ & $23.0\pm1.0$ & $0.32\pm0.06$ & $0.17\pm0.04$ & $0.15\pm0.03$\\
\qquad NGC\,2403 & \pt{0}$6.0\pm0.7$ & $13.4\pm1.0$ & $0.45\pm0.06$ & $0.35\pm0.10$ & $0.26\pm0.07$\\
\qquad NGC\,3077 & \pt{0}$5.9\pm0.2$ & $15.6\pm0.2$ & $0.38\pm0.02$ & $0.25\pm0.02$ & $0.20\pm0.01$\\
\qquad NGC\,3738 & $12.9\pm0.5$ & $30.9\pm1.5$ & $0.42\pm0.03$ & $0.89\pm0.13$ & $0.47\pm0.05$\\
\qquad NGC\,4449 & \pt{0}$7.4\pm1.0$ & $15.9\pm0.4$ & $0.47\pm0.06$ & $0.14\pm0.04$ & $0.13\pm0.04$\\
\qquad NGC\,5055 & \pt{0}$7.5\pm0.7$ & $17.3\pm0.9$ & $0.43\pm0.05$ & $0.38\pm0.09$ & $0.27\pm0.06$\\
\qquad NGC\,5236 & \pt{0}$5.7\pm0.5$ & $14.0\pm0.5$ & $0.41\pm0.04$ & $0.26\pm0.04$ & $0.21\pm0.03$\\
\qquad NGC\,5457 & $12.2\pm2.8$ & $12.2\pm4.2$ & $1.00\pm0.41$ & $2.33\pm0.69$ & $0.70\pm0.32$\\
\qquad NGC\,6946 & \pt{0}$6.0\pm0.2$ & $16.4\pm0.5$ & $0.37\pm0.02$ & $0.55\pm0.05$ & $0.35\pm0.03$\\
\qquad UGC\,8508 & \pt{0}$3.2\pm0.9$ & $11.1\pm0.2$ & $0.29\pm0.09$ & $0.04\pm0.01$ & $0.04\pm0.01$\\
\bottomrule
\end{tabular}
\vspace{2mm}
\end{table*}

Following the method described in Section~\ref{sec:s2}, we derive \HI super-profiles of the sample galaxies. As a comparison, we also derive two additional \HI super-profiles for each galaxy following the method described in \citetalias{Ianja+12}. As one of them, \citetalias{Ianja+12} use velocity profiles whose shapes are symmetric in velocity with respect to their centroid velocities when constructing an \HI super-profile. We call this {\sc sp}-based super-profile. To select symmetric velocity profiles, \citetalias{Ianja+12} exclude velocity profiles if the velocity difference between the centroid velocities derived using the {\sc moment1} and Hermite $h_3$ polynomial fitting analyses is greater than 2~\kms. Recently, several works (e.g., \citealt{Stilp+13, Ianja+15}) use central velocities of velocity profiles derived from \mbox{Hermite} $h_3$ fitting when constructing \HI super-profiles. We also construct \HI super-profiles using this method and call the resulting profiles {\sc her3}-based super-profiles. We present the \HI super-profiles of the sample galaxies in Appendix~\ref{sec:appA}.

We find that 17 galaxies show significant negative bowls (i.e., residuals in the wing parts greater than those in the center) in their \HI super-profiles constructed using the method in this paper (see DDO\,52, DDO\,87, DDO\,101, DDO\,165, DDO\,168, LGS\,3, NGC\,628, NGC\,1569, NGC\,2403, NGC\,3077, NGC\,3738, NGC\,4449, NGC\,5055, NGC\,5236, NGC\,5457, NGC\,6946, and UGC\,8508 in Figure~\ref{fig:2GFIT_super-profiles}). As discussed in \citetalias{Ianja+12}, this could be due to the short spacing problem of radio interferometries by which large scale emission is missed. It can be corrected by fitting a polynomial function to the wing parts which show significant fluctuations and subtracting the fit from the original super-profile as done in \citetalias{Ianja+12}. Alternatively, the negative bowl feature in the super-profile could be also caused by \HI absorptions, telescope defects or calibration issues. In contrast to the short spacing issue, these only affect specific parts of the spectra for which the correction with a polynomial function fitting is not appropriate. We visually inspected velocity profiles of the \HI data cubes of the 17 galaxies but were not able to clearly figure out what causes the negative bowl feature in their super-profiles. Thus, in this work, we omit the 17 galaxies in the \HI super-profile analysis. But see Appendix~\ref{sec:appB} as we also show the results with negative bowl corrected sample following the method described in \citetalias{Ianja+12}.

As described in Section~\ref{sec:s23}, we fit a double Gaussian model to the \HI super-profiles of the sample \mbox{galaxies}, and quantify their shapes. The resulting \HI super-profile parameters are given in Table~\ref{tab:GFIT_super-profiles}.

\subsection{Correlations}

\begin{figure*}[!p]
\centering
\includegraphics[width=\linewidth, trim={0mm, 5mm, 0mm, 2mm}, clip]{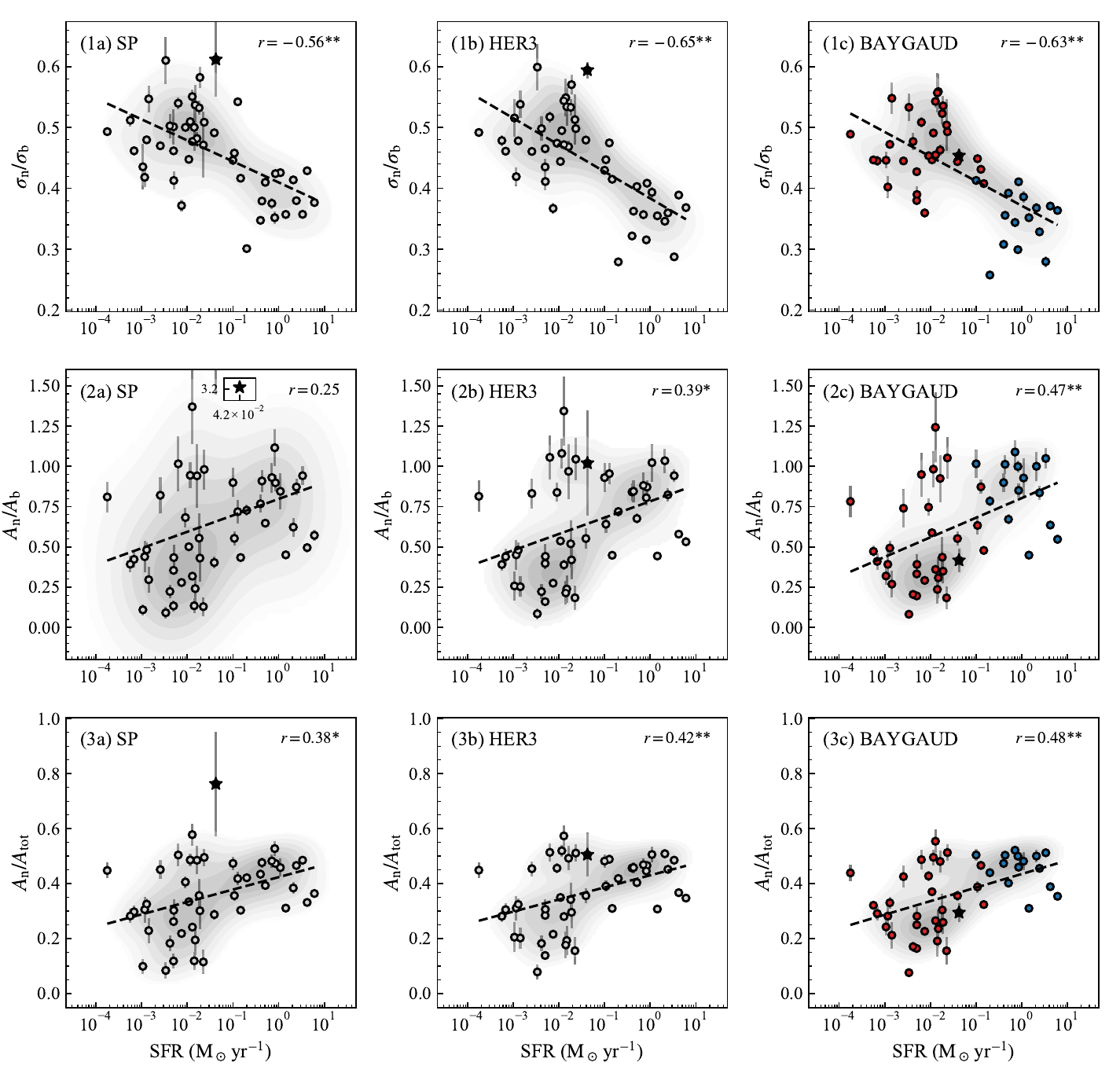}
\caption{Correlation between super-profile parameters and global star formation rate. Row (1): \snsb-SFR; the relations between the velocity dispersion ratio of the narrow and broad Gaussian components and SFR, (2): \anab-SFR; the relations between the area ratio of the narrow and broad Gaussian components, and (3): \anat-SFR; the relations between the area ratio of the narrow component and the total area (\at\ =\ \an\ +\ \ab,\ \anat). Column (a): ones from {\sc sp}-based super-profile, (b): ones from {\sc her3}-based super-profile, and (c): ones produced with our method. We show the Pearson correlation coefficient on the top right of each panel with its statistical significance (**: $p < 0.005$, *: $p < 0.05$). The dashed line in each panel shows global linear trend fitted with least-square method. In panel (2a), an outlier positioned outside the panel is shown in the inset. The markers with a filled star symbol are Haro\,36, which show the most significant change between different methods. The red and blue colored markers in column (c) represent irregular and spiral type galaxies, respectively.\label{fig:correlation}}
\vspace{2mm}
\end{figure*}

We investigate the correlations between the star formation rates of the sample galaxies and their \HI super-profile parameters. In this analysis, the 17 galaxies showing the negative bowl feature in their super-profiles are excluded as discussed in Section~\ref{sec:s42}. The correlations are presented in Figure~\ref{fig:correlation}; {\bf Top panels:} \snsb--SFR; the relations between the velocity dispersion ratio of the narrow ($\sigma_{\rm{n}}$) and broad ($\sigma_{\rm{b}}$) Gaussian components and SFR for the {\sc sp}, {\sc her3}, and {\sc baygaud}-based \HI super-profiles; {\bf Middle panels:} \anab--SFR; the relations between the area ratio of the narrow (\an) and broad (\ab) Gaussian components and SFR for the {\sc sp}, {\sc her3}, and {\sc baygaud}-based \HI super-profiles; {\bf Lower panels:} \anat--SFR; the relations between the area ratio of the narrow Gaussian component (\an) and the total area (\at\,=\,\an\,+\,\ab) and SFR for the {\sc sp}, {\sc her3}, and {\sc baygaud}-based \HI super-profiles; The Pearson correlation coefficients $r$ (**: $p<0.005$, *: $p<0.05$) derived for the relations are denoted in the top-right corner of each panel of Figure~\ref{fig:correlation}.

Despite the scatter, global trends of decreasing \snsb\ and increasing \anab\ and \anat\ with SFR are seen from the three \HI super-profile analyses. This is also found in \citetalias{Ianja+12}. The kinematically narrow \HI gas components are likely to be associated with star formation in galaxies. The higher SFRs are found in the galaxies which have larger fractions of kinematically narrow \HI gas with lower velocity dispersions. 

Compared to the {\sc sp}-based \HI super-profiles, both the {\sc her3} and {\sc baygaud}-based ones show smaller scatter in the correlations. This is also quantified by the Pearson correlation coefficients, indicating more negative (\snsb--SFR) and positive (\anab--SFR and \anat--SFR) correlations for the {\sc her3} and {\sc baygaud}-based super-profiles than the {\sc sp} ones. This is mainly caused by the exclusion of asymmetric velocity profiles in the {\sc sp}-based \HI super-profiles. As discussed earlier, star formation in galaxies gives rise to turbulent gas motions, which results in asymmetric velocity profiles. The contribution of these asymmetric velocity profiles which are associated with star formation in galaxies to the {\sc sp}-based \HI super-profiles could be relatively smaller than the {\sc her3} and {\sc baygaud}-based super-profiles. This effect would be particularly significant in galaxies with high SFRs (see the panel 2a in Figure~\ref{fig:correlation}).

The correlation results between the {\sc her3} and {\sc baygaud}-based \HI super-profiles are well consistent although the ones derived from {\sc baygaud}-based ones show slightly smaller scatter and stronger correlations. As discussed earlier, the {\sc baygaud}-based \HI super-profiles would be much superior to the {\sc her3}-based ones for extracting kinematically narrower \HI gas from velocity profiles which have additional kinematic components with significant amplitudes. The fractions of these heavily disturbed velocity profiles may not be high in our sample galaxies. The {\sc baygaud}-based \HI super-profiles are also expected to be less affected by beam smearing. The effect of beam smearing on the \HI super-profiles would be more visible in low-resolution data. Therefore, the difference between the {\sc her3} and {\sc baygaud}-based \HI super-profiles may not be clearly visible in the high-resolution \HI data of both THINGS and LITTLE THINGS where the beam smearing effect is highly reduced. In this respect, use of low-resolution \HI data cubes of galaxies would be \mbox{interesting} for \mbox{further} testing the {\sc baygaud}-based \HI super-profiles.

\begin{figure}[!t]
\centering
\includegraphics[width=\linewidth]{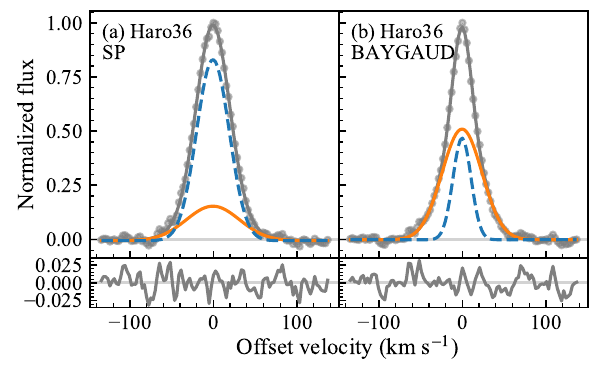}
\caption{\HI super-profiles of Haro\,36 built with different methods. Panel (a): the {\sc sp} super-profile, and (b): the {\sc baygaud} super-profile. The peak of each super-profile is normalized to compare the shapes.}
\label{fig:haro36_compare}
\vspace{2mm}
\end{figure}

There are galaxies whose \HI super-profiles derived using the three methods show significant differences in the correlations. As an example, we show a galaxy, Haro\,36 which is marked with black filled star symbol in Figure~\ref{fig:correlation}. We compare its {\sc sp} and {\sc baygaud}-based \HI super-profiles in Figure~\ref{fig:haro36_compare}. For the comparison, we normalize the profiles with their peak fluxes. The {\sc sp}-based super-profile has a blunter peak compared to the {\sc baygaud}-based one. This is mainly because of the profile decomposition based on which the {\sc baygaud}-based super-profile is constructed. Stacking of raw velocity profiles with multiple kinematic components but not decomposed would result in such a blunter profile. As shown in Figure~\ref{fig:haro36_compare}, the {\sc baygaud}-based \HI super-profile with a relatively narrower core and a broader wing has higher fraction of the kinematically narrow Gaussian component in the {\sc 2gfit} parameterization than the {\sc sp}-based profile. Despite the high-resolution LITTLE THINGS observations, low mass dwarf galaxies like Haro\,36 could be relatively more affected by beam smearing than the other galaxies due to the small size of its gas disk (e.g., $M_{\rm HI}$-size relation in \citealt{Wang+16}). As shown in Figure~\ref{fig:correlation}, the correlations for Haro\,36 (black filled dots) derived using the shape parameters of the {\sc baygaud}-based \HI superprofie are much better consistent with the global relations indicated by the dashed lines than the others (i.e., {\sc sp} and {\sc her3}-based super-profiles).

The red and blue colored markers in panels (1c, 2c and 3c) in Figure~\ref{fig:correlation} represent irregular and spiral galaxies, respectively\footnote{Galaxy type information from {\sc simbad} (\url{http://simbad.u-strasbg.fr/simbad/}, \citealt{Wenger+00})}. Compared to irregular type galaxies, spiral galaxies lie in the high-SFR end, following the linear extension of the global trend shown in each panel. This may provide additional information on classifying galaxy types in high-z observations.

\section{Summary\label{sec:s5}}

In this paper, we present a new method which constructs an \HI super-profile of a galaxy by stacking velocity profiles in a data cube. This S/N improved super-profile is useful for deriving the galaxy’s global \HI properties like velocity dispersion and mass from observations which do not provide sufficient surface brightness sensitivity for the galaxy. The main difference between ours and other super-profile methods (\citetalias{Ianja+12}, \citealt{Stilp+13}) is that the new method first decomposes individual velocity profiles of the cube with an optimal number of Gaussian components using a profile decomposition tool, {\sc baygaud}. It then aligns the decomposed velocity profiles in velocity using their centroid velocities which are determined from the Gaussian fits. The \HI super-profile is constructed by co-adding all the aligned profiles. This is compared to the other methods where the original velocity profiles of the cube are aligned and co-added using their centroid velocities determined from the {\sc moment1} or Hermite $h_3$ polynomial fitting analyses.

The previous methods have limitations in estimating the centroid velocities of asymmetric velocity profiles with multiple kinematic components, and their resulting \HI super-profiles become blunter in shape with a broader core. On the other hand, the so-called {\sc baygaud}-based \HI super-profile is able to take even highly asymmetric velocity profiles with multiple kinematic components into account via the profile decomposition analysis.

We make a practical application of the new method to a sample of nearby galaxies from THINGS and LITTLE THINGS to construct their \HI superprofies. We use the high-resolution THINGS and LITTLE THINGS \HI data cubes of the sample galaxies. In addition, for the comparison between the new and previous methods, we also construct two additional \HI super-profiles of the sample galaxies using symmetric and all velocity profiles whose centroid velocities are determined from Hermite $h_3$ polynomial fitting, respectively. These are called {\sc sp} and {\sc her3}-based \HI super-profiles in this work.

In general, the {\sc baygaud}-based \HI super-profiles of the sample galaxies have narrower cores and broader wings in shape than the other two super-profiles. This is because the {\sc baygaud}-based super-profile co-adds more kinematically narrow velocity profiles being deblended from the original asymmetric profiles. We fit a double Gaussian model to the \HI super-profiles to quantify their shapes. The shape parameters (\snsb, \anab, and \anat) are then correlated with the SFRs of the sample galaxies. We find strong correlations that \snsb\ decreases but \anab\ and \anat\ increase as the SFRs increase, respectively. This is also found in \citetalias{Ianja+12}, and indicates that the kinematically narrower Gaussian components of \HI super-profiles are closely associated with star formation and/or stellar feedback in the galaxies.

The relatively weaker correlations with the larger scatter for the {\sc sp}-based \HI super-profiles than the others are mainly attributed to the use of symmetric velocity profiles. The intentional exclusion of asymmetric velocity profiles which could be caused by star formation in the galaxies results in the weaker correlations with larger scatter. The correlations derived using the {\sc her3}-based and {\sc baygaud}-based super-profiles are comparable but the {\sc baygaud}-based super-profile analysis gives slightly smaller scatter. It is found that the scatter is much reduced in the {\sc baygaud}-based super-profile correlations for galaxies with low mass whose \HI disk size is small. These relatively small galaxies are likely to be affected by beam smearing which is not corrected in the conventional super-profile method. On the other hand, the effect of beam smearing is reduced in the {\sc baygaud}-based \HI super-profiles constructed using the decomposed profiles. In this regard, further tests of {\sc baygaud}-based \HI super-profile analysis using low or intermediate resolution \HI data cubes of galaxies which are severely affected by beam smearing would be interesting for future work.


\acknowledgments

S.-H. Oh acknowledges a support from the National Research Foundation of Korea (NRF) grant funded by the Korea government (Ministry of Science and ICT: MSIT) (No. NRF-2020R1A2C1008706).



\newpage

\appendix

\onecolumn
\begingroup

\counterwithin{figure}{section}
\counterwithin{table}{section}

\section{2gfit Parameterization Results of Super-Profiles\label{sec:appA}}

\centerline{In this appendix, we show {\sc 2gfit} parameterization results of our sample in Figure~\ref{fig:2GFIT_super-profiles}.}

\begin{figure*}[!h]
\centering
\captionsetup[subfigure]{labelformat=empty}
\subfloat[]{\includegraphics[width=0.27\linewidth, trim={0mm 0mm 0mm 3mm}, clip]{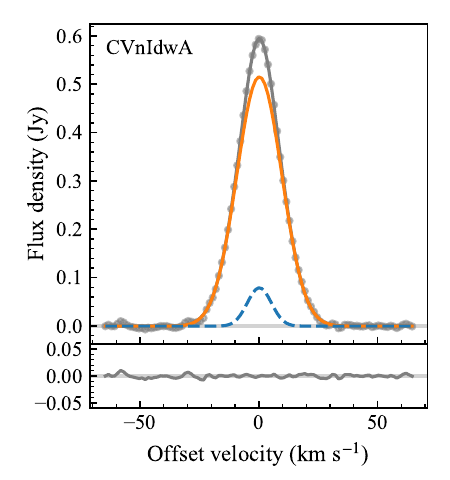}}
\subfloat[]{\includegraphics[width=0.27\linewidth, trim={0mm 0mm 0mm 3mm}, clip]{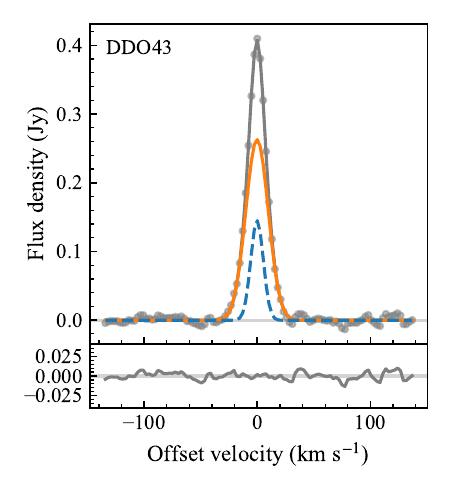}}
\subfloat[]{\includegraphics[width=0.27\linewidth, trim={0mm 0mm 0mm 3mm}, clip]{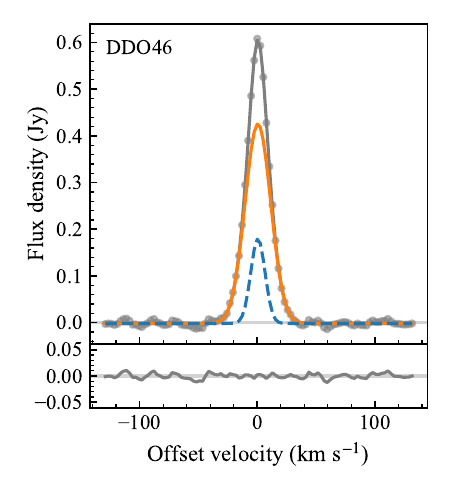}}\\[-5ex]
\subfloat[]{\includegraphics[width=0.27\linewidth]{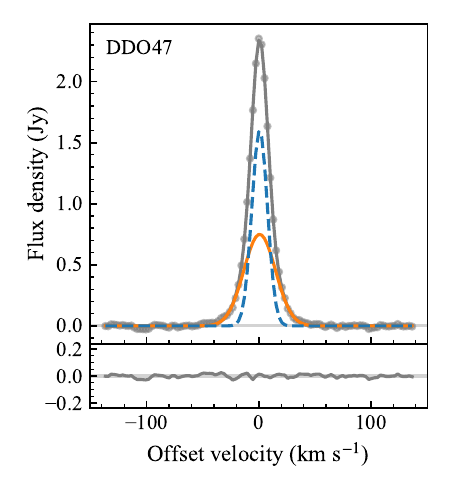}}
\subfloat[]{\includegraphics[width=0.27\linewidth]{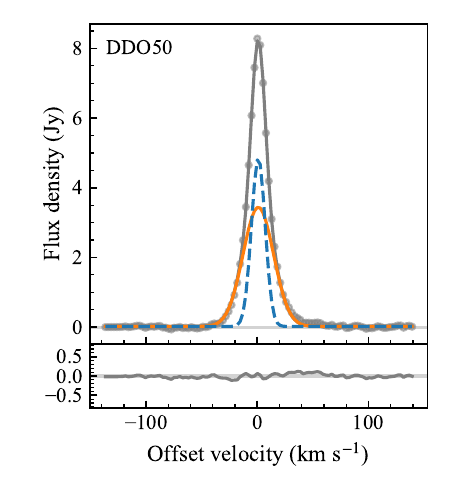}}
\subfloat[]{\includegraphics[width=0.27\linewidth]{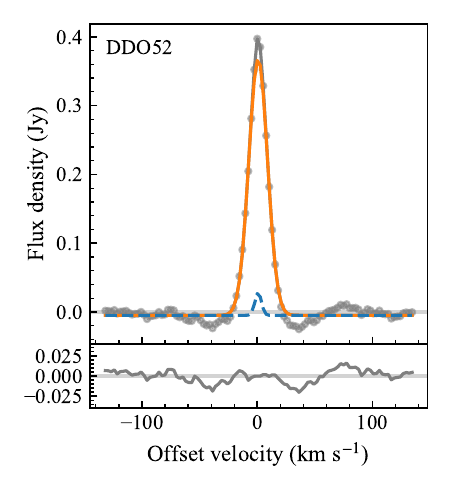}}\\[-5ex]
\subfloat[]{\includegraphics[width=0.27\linewidth]{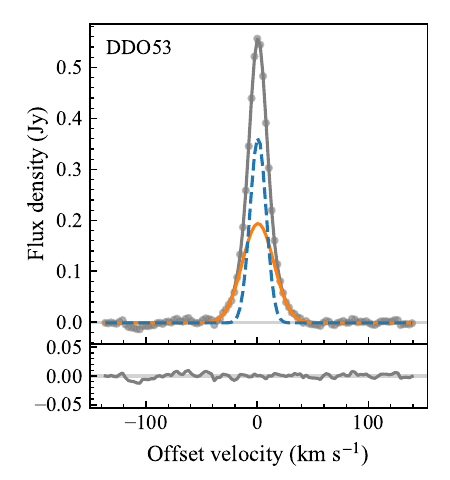}}
\subfloat[]{\includegraphics[width=0.27\linewidth]{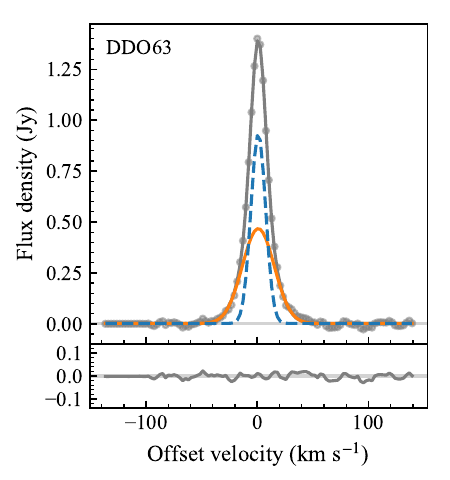}}
\subfloat[]{\includegraphics[width=0.27\linewidth]{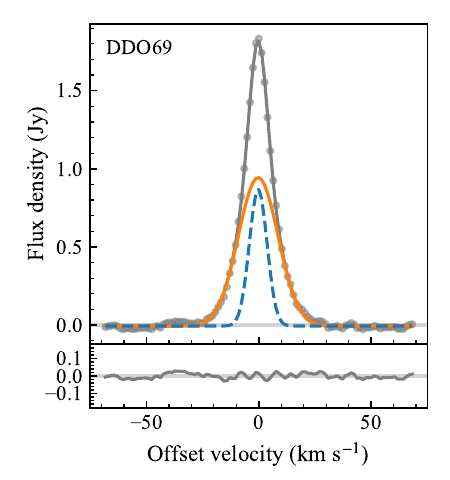}}\\[-5ex]
\subfloat[]{\includegraphics[width=0.27\linewidth]{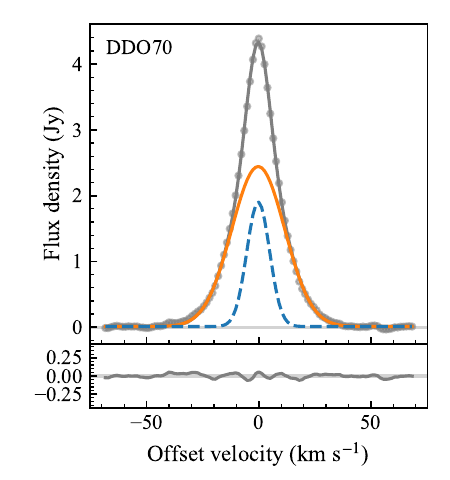}}
\subfloat[]{\includegraphics[width=0.27\linewidth]{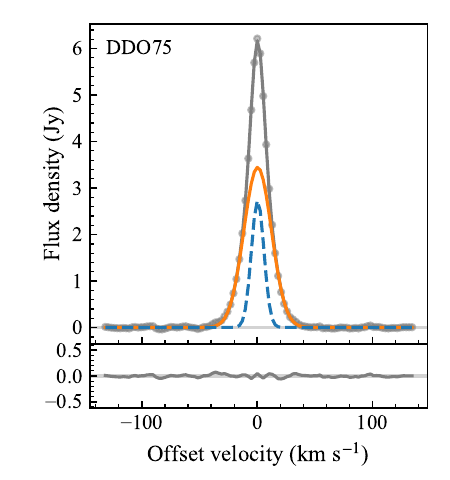}}
\subfloat[]{\includegraphics[width=0.27\linewidth]{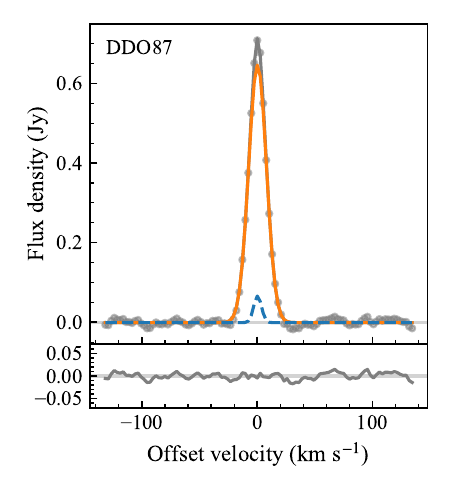}}\\
\vspace{-3mm}
\caption{Super-profiles with {\sc 2gfit} parameterization of our sample galaxies. Grey circles with bars indicate data points with 3-$\sigma$ uncertainties although they are generally small than the markers. Blue dashed line and orange solid line show narrow and broad component, respectively. The solid grey line shows the sum of narrow and broad components. On the bottom of every panels, we show the residuals of corresponding fit with y-range set as $\pm$10\% of the super-profile's peak. }
\label{fig:2GFIT_super-profiles}
\end{figure*}

\begin{figure*}[!ph]
\ContinuedFloat
\centering
\captionsetup[subfigure]{labelformat=empty}
\subfloat[]{\includegraphics[width=0.28\linewidth]{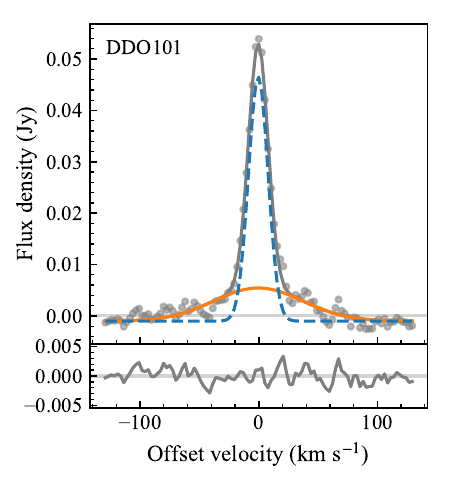}}
\subfloat[]{\includegraphics[width=0.28\linewidth]{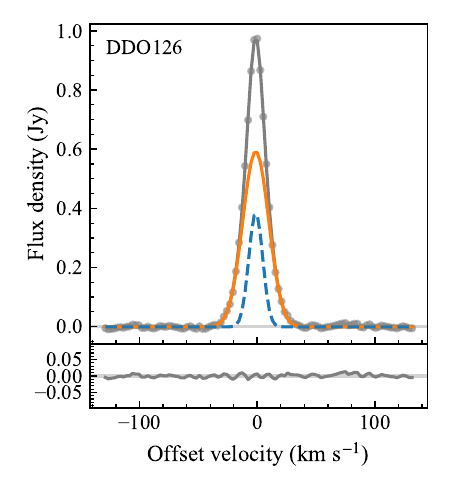}}
\subfloat[]{\includegraphics[width=0.28\linewidth]{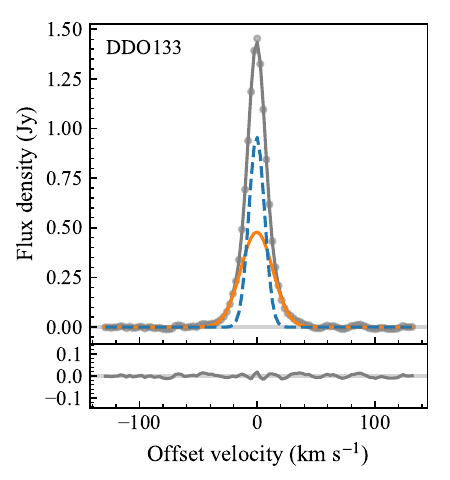}}\\[-5ex]
\subfloat[]{\includegraphics[width=0.28\linewidth]{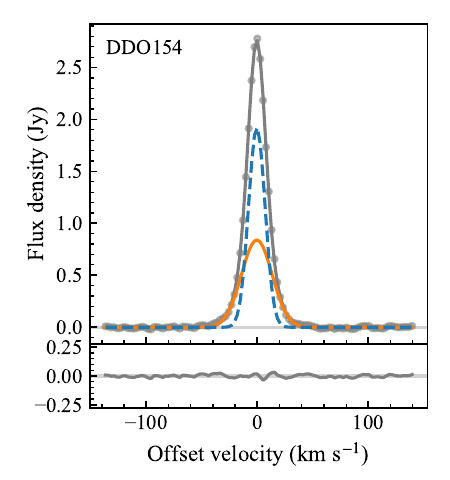}}
\subfloat[]{\includegraphics[width=0.28\linewidth]{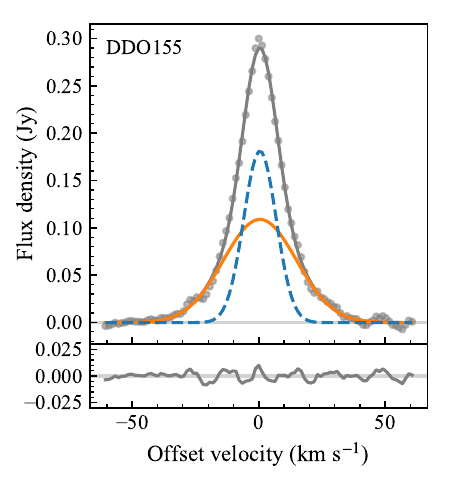}}
\subfloat[]{\includegraphics[width=0.28\linewidth]{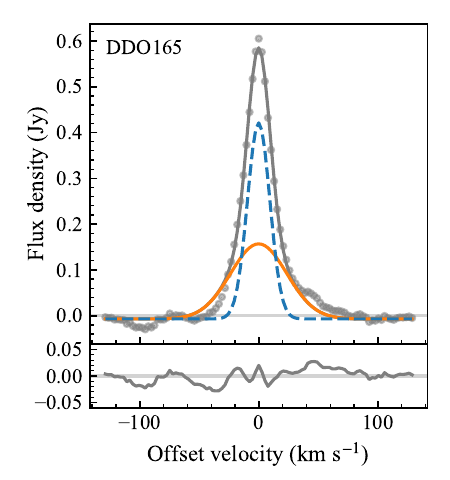}}\\[-5ex]
\subfloat[]{\includegraphics[width=0.28\linewidth]{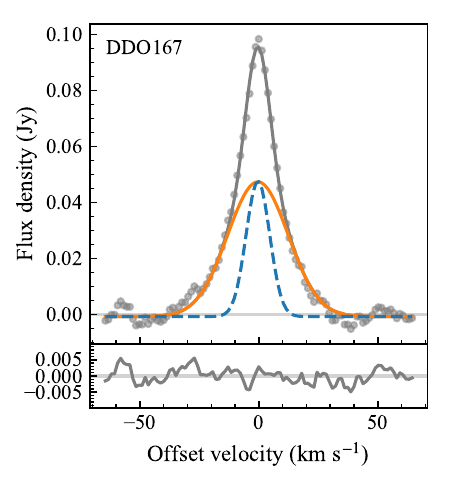}}
\subfloat[]{\includegraphics[width=0.28\linewidth]{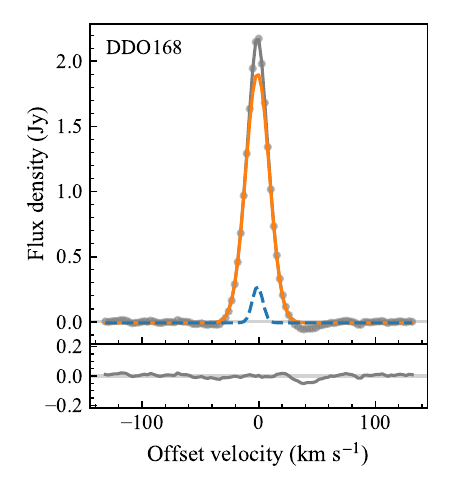}}
\subfloat[]{\includegraphics[width=0.28\linewidth]{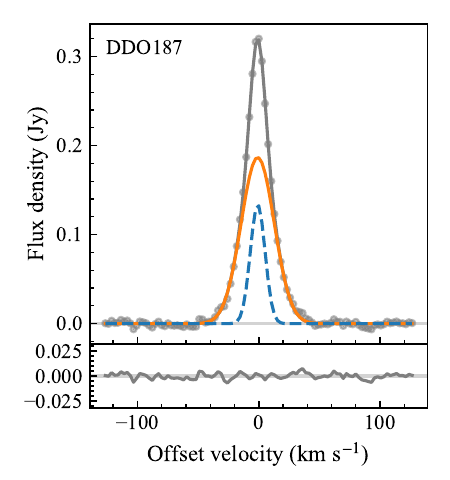}}\\[-5ex]
\subfloat[]{\includegraphics[width=0.28\linewidth]{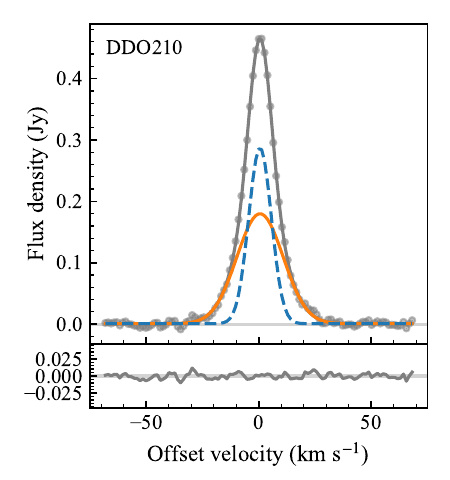}}
\subfloat[]{\includegraphics[width=0.28\linewidth]{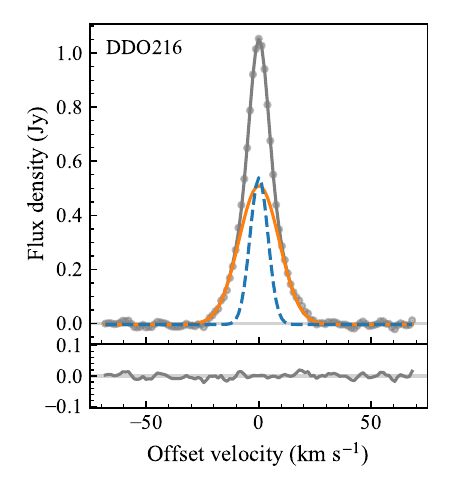}}
\subfloat[]{\includegraphics[width=0.28\linewidth]{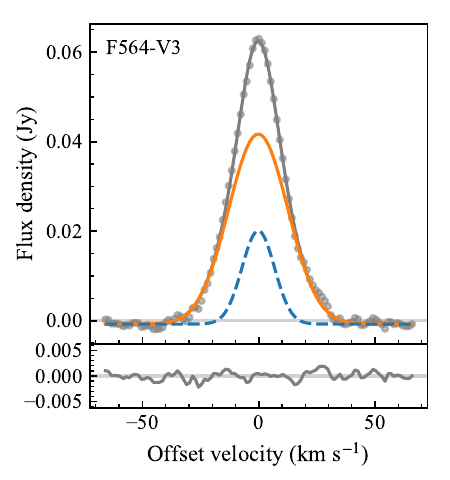}}\\
\vspace{-3mm}
\caption{Continued}
\end{figure*}

\begin{figure*}[!ph]
\ContinuedFloat
\centering
\captionsetup[subfigure]{labelformat=empty}
\subfloat[]{\includegraphics[width=0.28\linewidth]{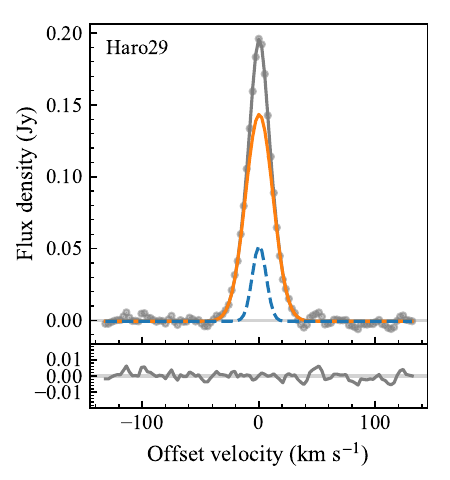}}
\subfloat[]{\includegraphics[width=0.28\linewidth]{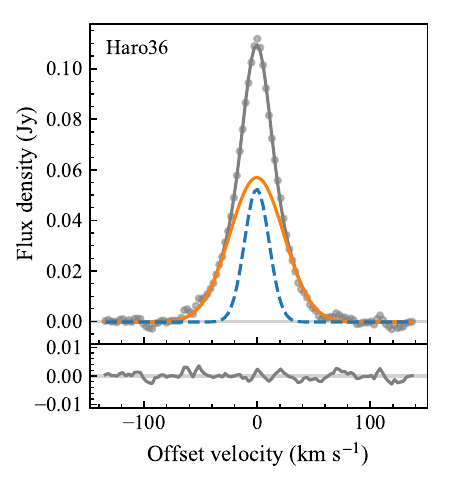}}
\subfloat[]{\includegraphics[width=0.28\linewidth]{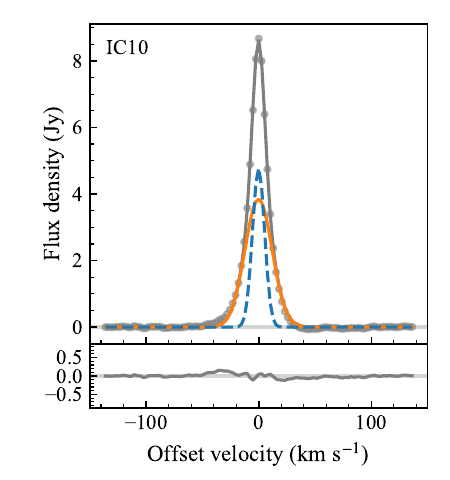}}\\[-5ex]
\subfloat[]{\includegraphics[width=0.28\linewidth]{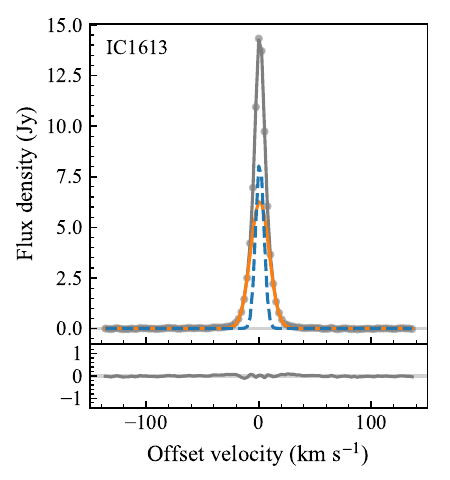}}
\subfloat[]{\includegraphics[width=0.28\linewidth]{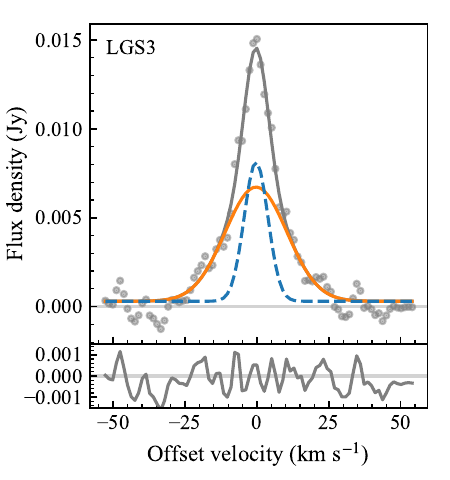}}
\subfloat[]{\includegraphics[width=0.28\linewidth]{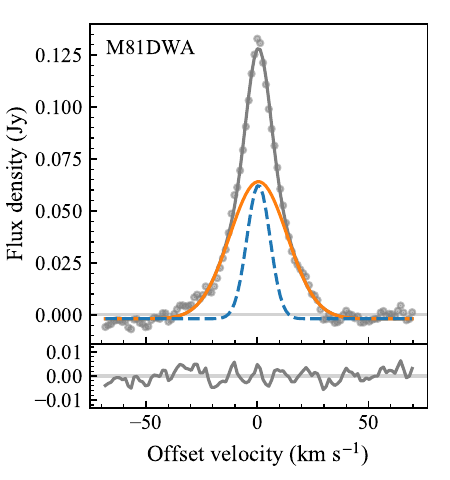}}\\[-5ex]
\subfloat[]{\includegraphics[width=0.28\linewidth]{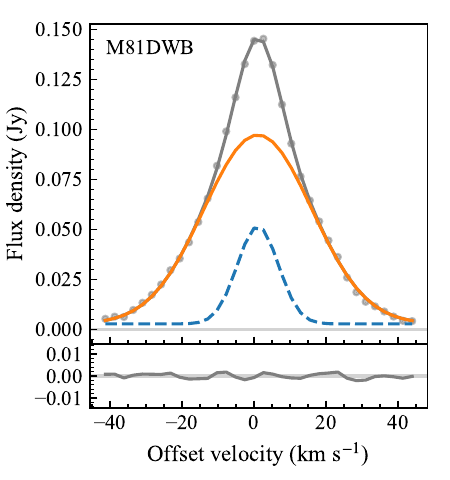}}
\subfloat[]{\includegraphics[width=0.28\linewidth]{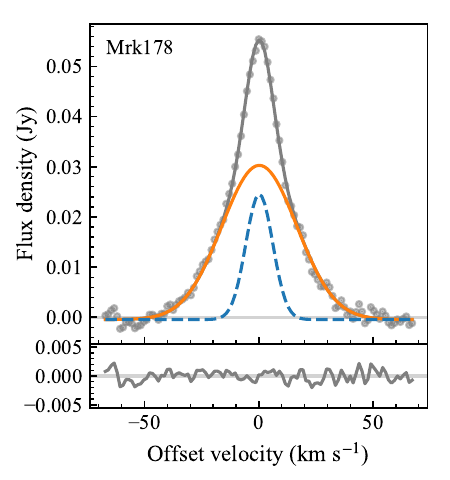}}
\subfloat[]{\includegraphics[width=0.28\linewidth]{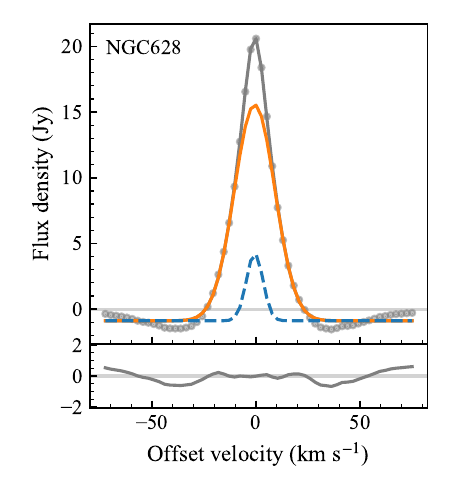}}\\[-5ex]
\subfloat[]{\includegraphics[width=0.28\linewidth]{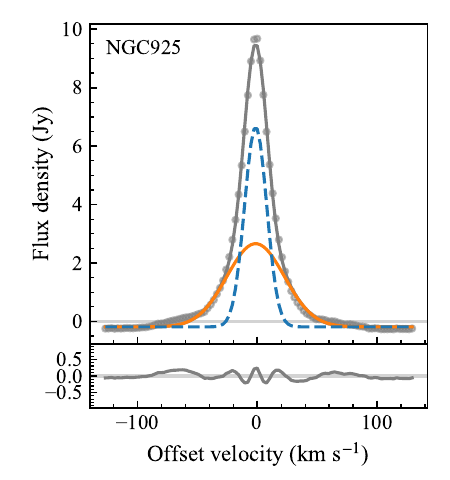}}
\subfloat[]{\includegraphics[width=0.28\linewidth]{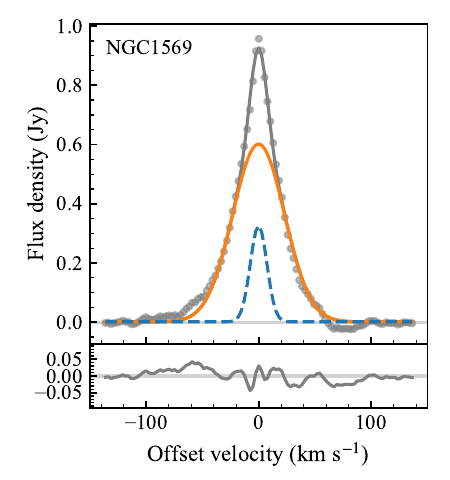}}
\subfloat[]{\includegraphics[width=0.28\linewidth]{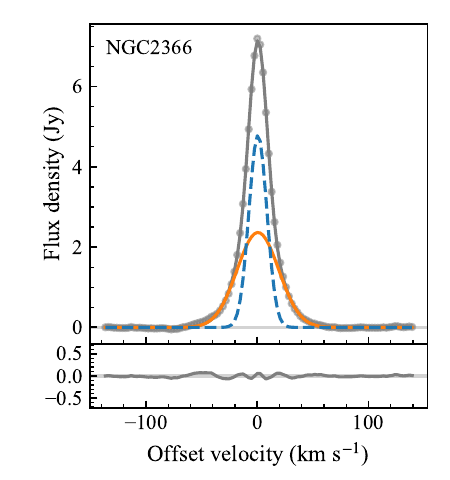}}\\
\vspace{-3mm}
\caption{Continued}
\end{figure*}

\begin{figure*}[!ph]
\ContinuedFloat
\centering
\captionsetup[subfigure]{labelformat=empty}
\subfloat[]{\includegraphics[width=0.28\linewidth]{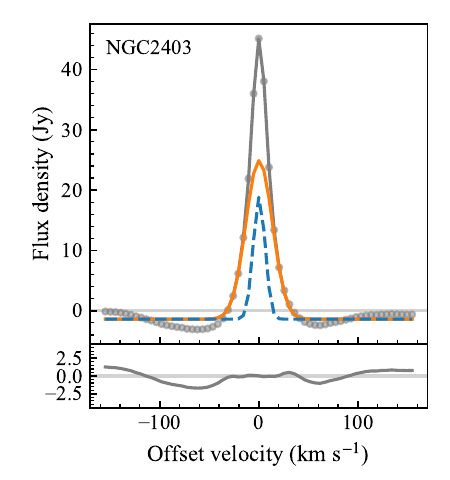}}
\subfloat[]{\includegraphics[width=0.28\linewidth]{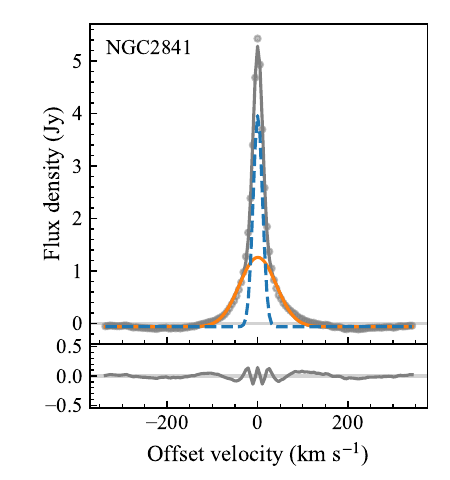}}
\subfloat[]{\includegraphics[width=0.28\linewidth]{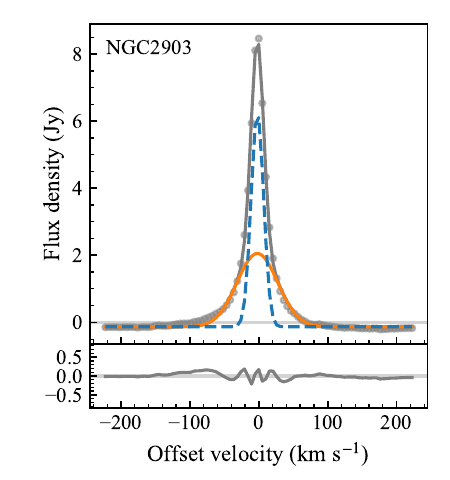}}\\[-5ex]
\subfloat[]{\includegraphics[width=0.28\linewidth]{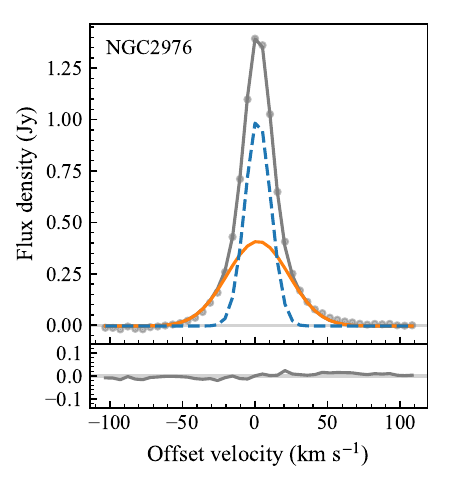}}
\subfloat[]{\includegraphics[width=0.28\linewidth]{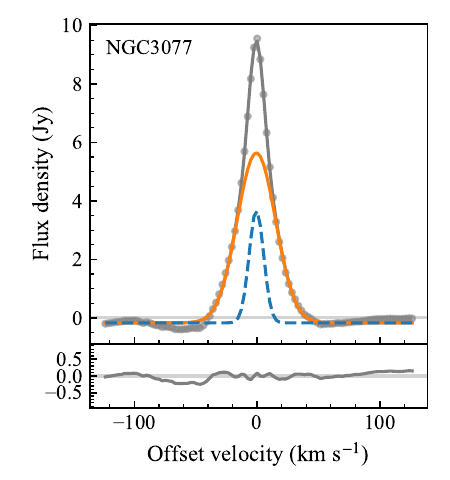}}
\subfloat[]{\includegraphics[width=0.28\linewidth]{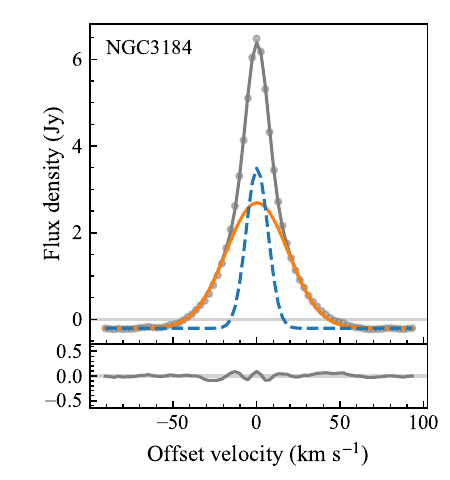}}\\[-5ex]
\subfloat[]{\includegraphics[width=0.28\linewidth]{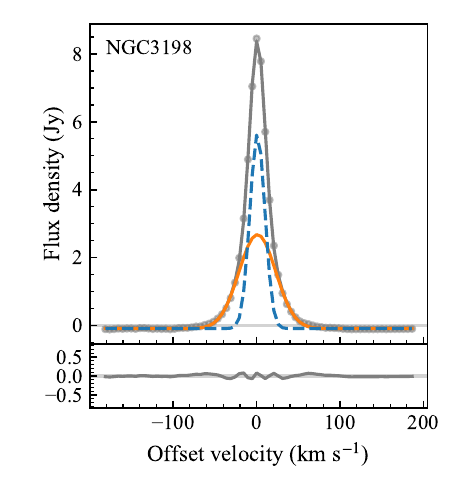}}
\subfloat[]{\includegraphics[width=0.28\linewidth]{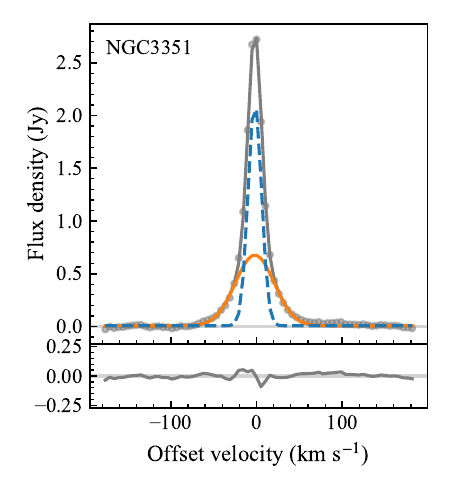}}
\subfloat[]{\includegraphics[width=0.28\linewidth]{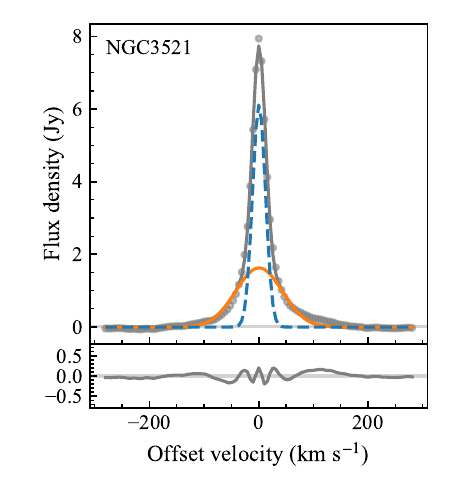}}\\[-5ex]
\subfloat[]{\includegraphics[width=0.28\linewidth]{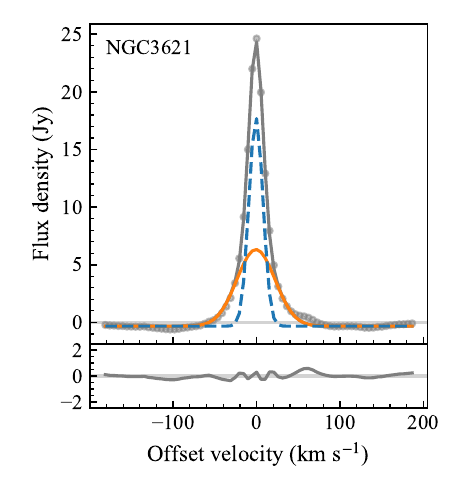}}
\subfloat[]{\includegraphics[width=0.28\linewidth]{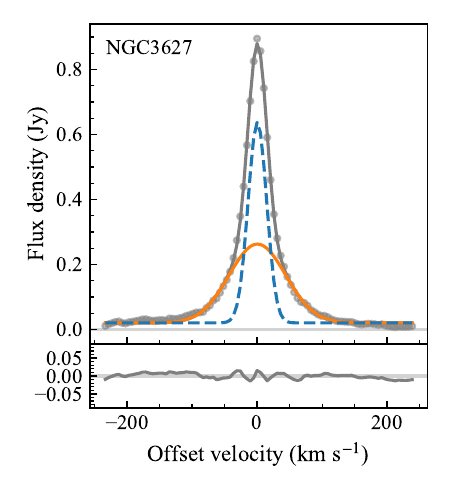}}
\subfloat[]{\includegraphics[width=0.28\linewidth]{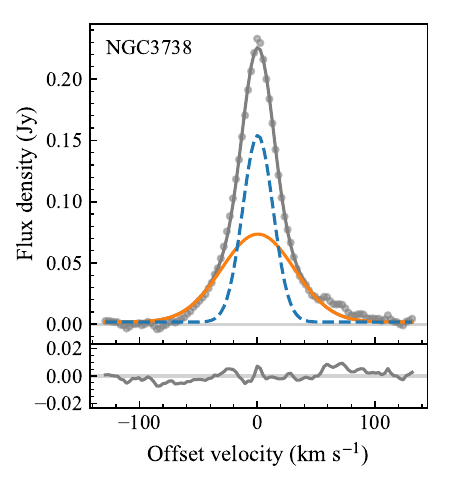}}\\
\vspace{-3mm}
\caption{Continued}
\end{figure*}

\begin{figure*}[!ph]
\ContinuedFloat
\centering
\captionsetup[subfigure]{labelformat=empty}
\subfloat[]{\includegraphics[width=0.28\linewidth]{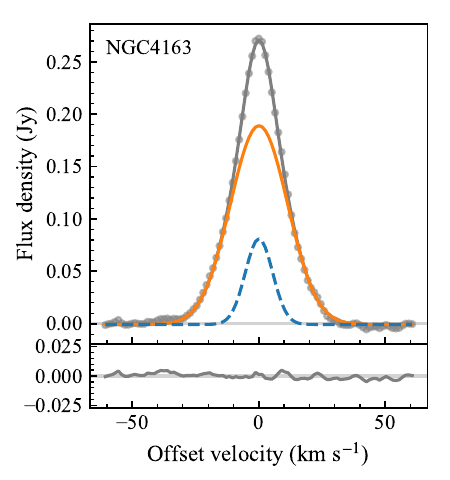}}
\subfloat[]{\includegraphics[width=0.28\linewidth]{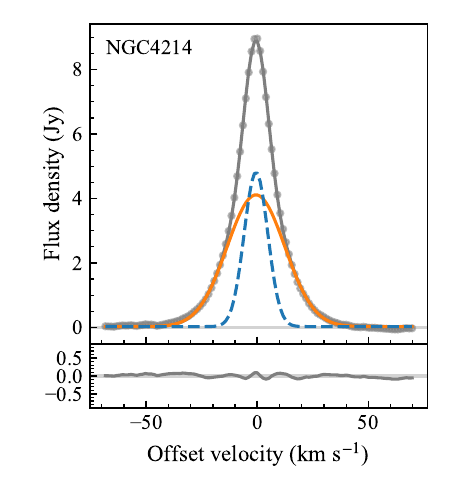}}
\subfloat[]{\includegraphics[width=0.28\linewidth]{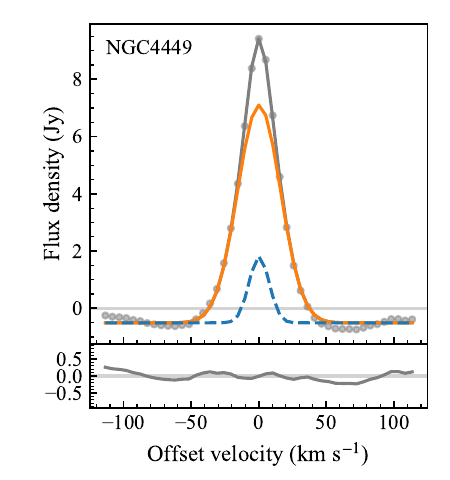}}\\[-5ex]
\subfloat[]{\includegraphics[width=0.28\linewidth]{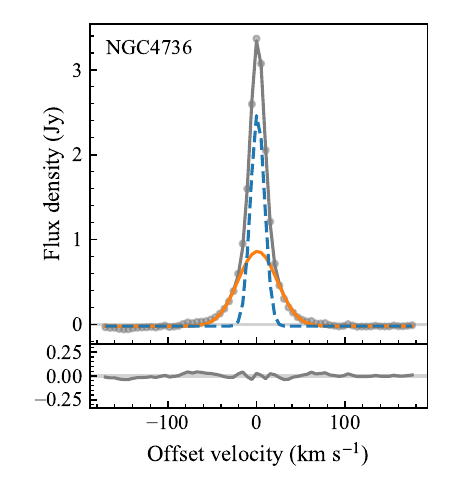}}
\subfloat[]{\includegraphics[width=0.28\linewidth]{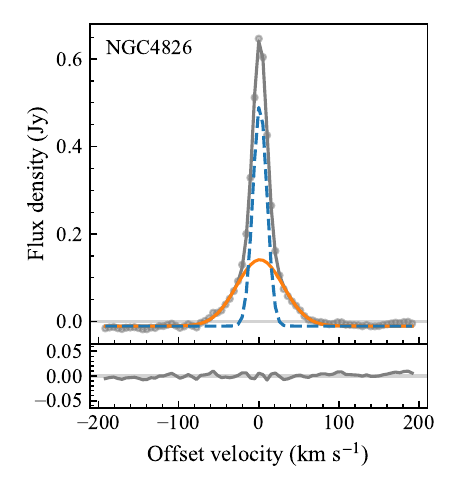}}
\subfloat[]{\includegraphics[width=0.28\linewidth]{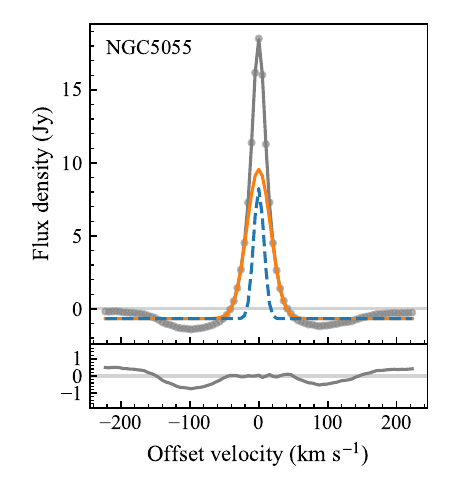}}\\[-5ex]
\subfloat[]{\includegraphics[width=0.28\linewidth]{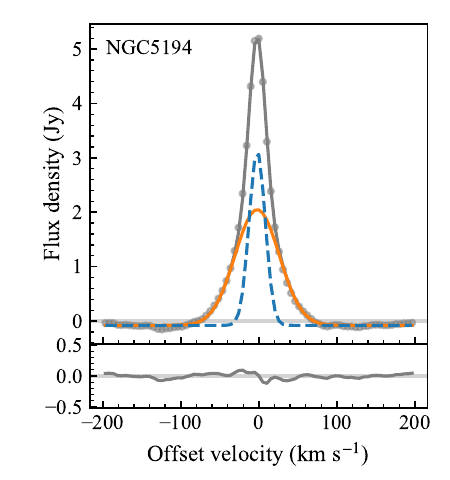}}
\subfloat[]{\includegraphics[width=0.28\linewidth]{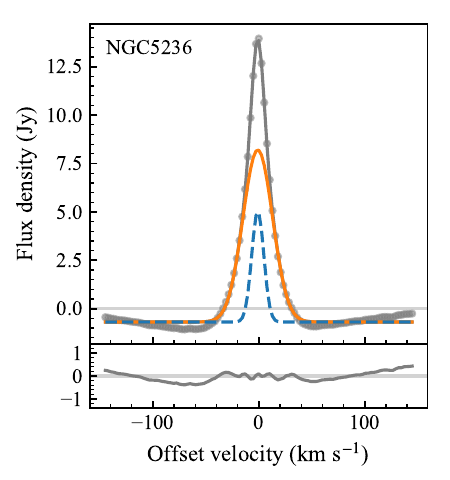}}
\subfloat[]{\includegraphics[width=0.28\linewidth]{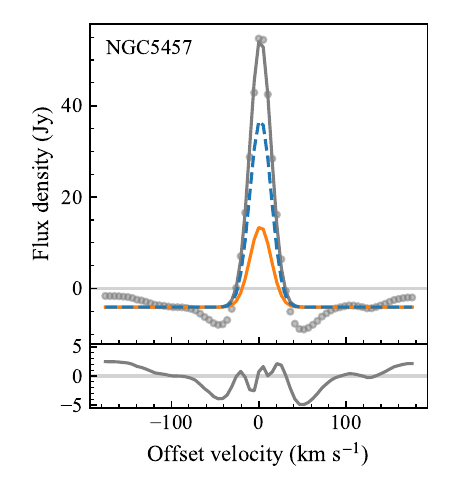}}\\[-5ex]
\subfloat[]{\includegraphics[width=0.28\linewidth]{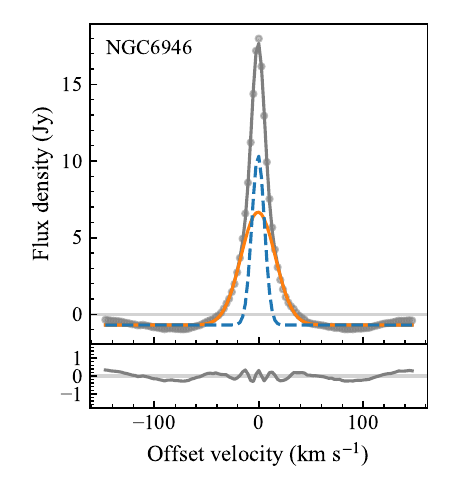}}
\subfloat[]{\includegraphics[width=0.28\linewidth]{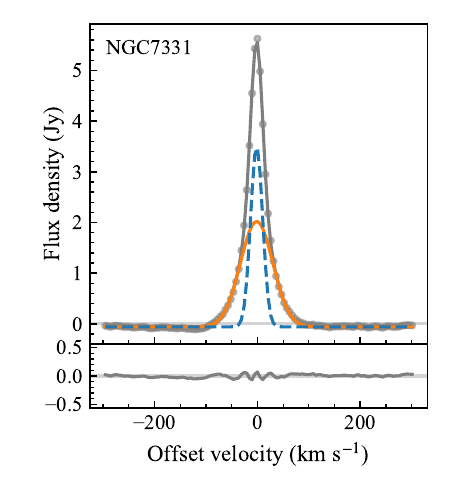}}
\subfloat[]{\includegraphics[width=0.28\linewidth]{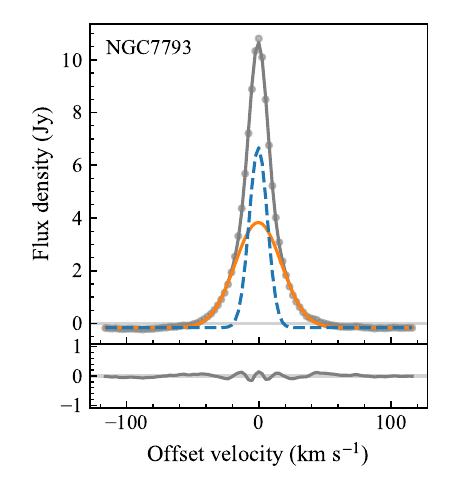}}\\
\vspace{-3mm}
\caption{Continued}
\end{figure*}

\begin{figure*}[!t]
\ContinuedFloat
\centering
\captionsetup[subfigure]{labelformat=empty}
\subfloat[]{\includegraphics[width=0.28\linewidth, trim={0mm 0mm 0mm 4mm}, clip]{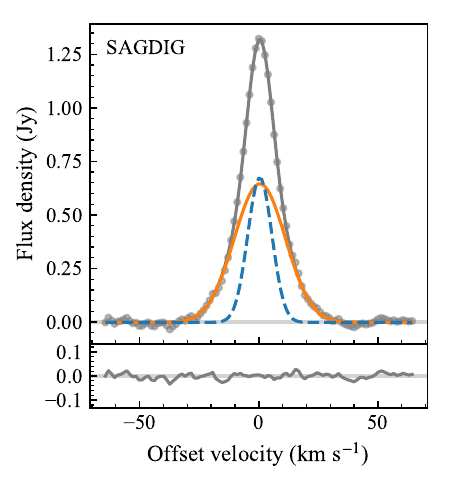}}
\subfloat[]{\includegraphics[width=0.28\linewidth, trim={0mm 0mm 0mm 4mm}, clip]{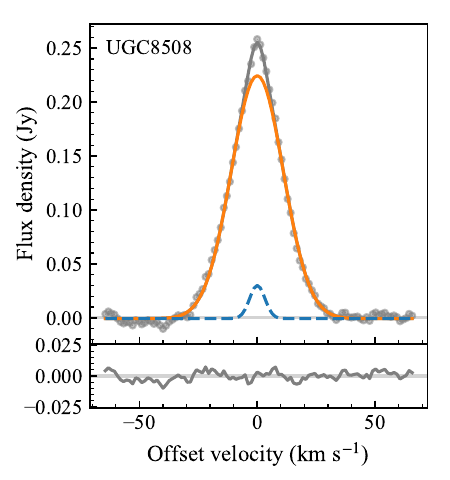}}
\subfloat[]{\includegraphics[width=0.28\linewidth, trim={0mm 0mm 0mm 4mm}, clip]{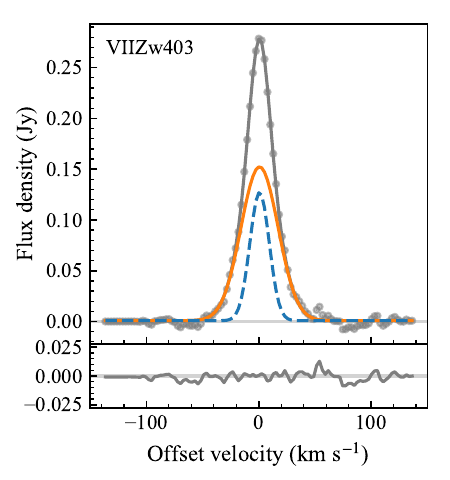}}\\[-3ex]
\subfloat[]{\includegraphics[width=0.28\linewidth]{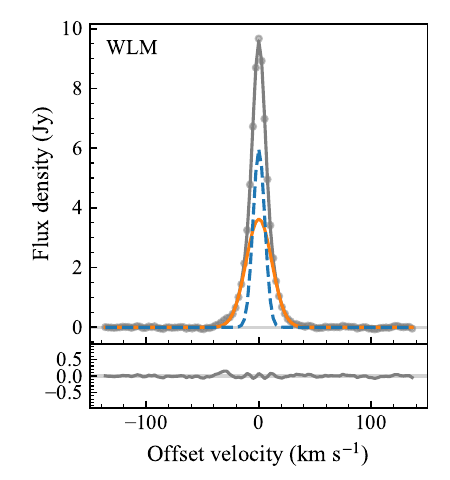}}\\
\vspace{-3mm}
\caption{Continued}
\vspace{3mm}
\end{figure*}

\newpage

\section{Negative-Bowl-Corrected Super-Profiles\label{sec:appB}}

We correct for negative bowls following the method described in \citetalias{Ianja+12}. That is done by fitting a polynomial to the wing part and subtracting from the original super-profile. Among 17 super-profiles showing negative bowl or defects as presented in Table~\ref{tab:GFIT_super-profiles}, we apply correction for six super-profiles (NGC\,628, NGC\,2403, NGC\,4449, NGC\,5055, NGC\,5236 and NGC\,6946) which show negative bowls build up gently towards the center without significant asymmetries. We present the {\sc 2gfit} result for these sample in Table~\ref{tab:GFIT_nbcorr} and in Figure~\ref{fig:GFIT_nbcorr}. In Figure~\ref{fig:correlation_nbcorr}, we show correlation including these negative bowl corrected sample.

\begin{table*}[!hb]
\caption{{\sc 2gfit} parameterization results for negative-bowl-corrected super-profile sample.\label{tab:GFIT_nbcorr}}
\centering\small
\begin{tabular}{x{0.166\linewidth} *{5}{y{0.166\linewidth}}}
\toprule
\qquad Galaxy & $\sigma_{\rm{n}}$ & $\sigma_{\rm{b}}$ & \snsb & \anab & \anat\\
       & [\kms] & [\kms] & & & \\
       & (1) & (2) & (3) & (4) & (5)\\
\midrule
\multicolumn{6}{c}{Before correction}\\
\midrule
\qquad NGC\,628\pt{0} & $4.0\pm0.7$ & ~$9.7\pm3.8$ & $0.41\pm0.03$ & $0.13\pm0.03$ & $0.11\pm0.02$ \\ 
\qquad NGC\,2403 & $6.0\pm0.6$ & $13.4\pm3.5$ & $0.45\pm0.02$ & $0.35\pm0.09$ & $0.26\pm0.05$ \\ 
\qquad NGC\,4449 & $7.4\pm1.1$ & $15.9\pm0.5$ & $0.47\pm0.04$ & $0.14\pm0.04$ & $0.13\pm0.03$ \\ 
\qquad NGC\,5055 & $7.5\pm0.7$ & $17.3\pm0.9$ & $0.43\pm0.02$ & $0.38\pm0.09$ & $0.27\pm0.05$ \\ 
\qquad NGC\,5236 & $5.7\pm0.4$ & $14.0\pm0.3$ & $0.41\pm0.02$ & $0.26\pm0.04$ & $0.21\pm0.02$ \\ 
\qquad NGC\,6946 & $6.0\pm0.2$ & $16.4\pm0.5$ & $0.37\pm0.01$ & $0.55\pm0.05$ & $0.35\pm0.02$ \\ 
\midrule
\multicolumn{6}{c}{After correction}\\
\midrule
\qquad NGC\,628\pt{0} & $4.9\pm0.3$ & $11.7\pm3.6$ & $0.42\pm0.13$ & $0.25\pm0.08$ & $0.20\pm0.05$\\
\qquad NGC\,2403 & $7.2\pm0.6$ & $18.1\pm0.8$ & $0.40\pm0.05$ & $0.61\pm0.08$ & $0.38\pm0.06$\\
\qquad NGC\,4449 & $9.9\pm0.3$ & $20.0\pm0.4$ & $0.49\pm0.02$ & $0.45\pm0.04$ & $0.31\pm0.03$\\
\qquad NGC\,5055 & $9.8\pm0.2$ & $28.5\pm1.0$ & $0.34\pm0.01$ & $0.79\pm0.06$ & $0.44\pm0.02$\\
\qquad NGC\,5236 & $6.6\pm0.3$ & $16.4\pm0.3$ & $0.40\pm0.02$ & $0.40\pm0.03$ & $0.28\pm0.02$\\
\qquad NGC\,6946 & $6.9\pm0.2$ & $21.7\pm0.9$ & $0.32\pm0.02$ & $0.73\pm0.09$ & $0.42\pm0.04$\\
\bottomrule
\end{tabular}
\tabnote{\textbf{Notes.} (1) Velocity dispersion of narrow Gaussian component; (2) Velocity dispersion of broad Gaussian component; (3) Ratio of velocity dispersion of the narrow and broad Gaussian component; (4) Ratio of area of narrow and broad Gaussian component; (5) Ratio of area of narrow Gaussian component and the total area (\at=\an+\ab).}
\vspace{-10mm}
\end{table*}

\begin{figure*}[!p]
\centering
\captionsetup[subfigure]{labelformat=empty}
\subfloat[]{\includegraphics[width=0.5\linewidth]{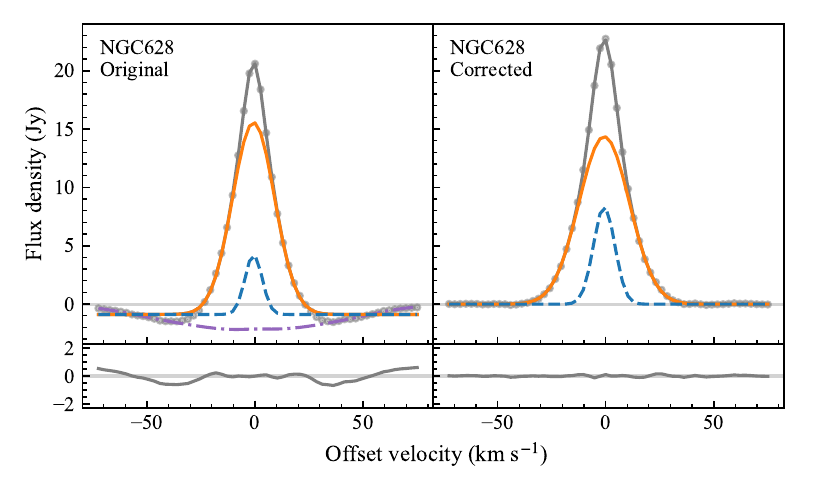}}
\subfloat[]{\includegraphics[width=0.5\linewidth]{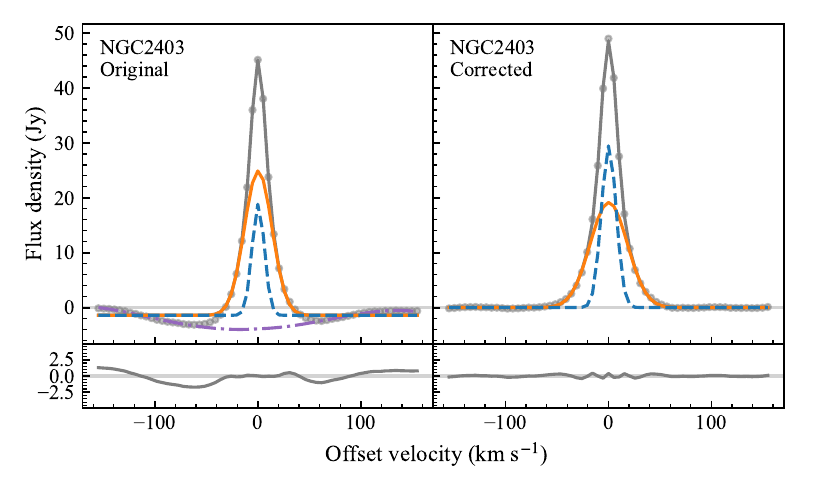}}\\
\subfloat[]{\includegraphics[width=0.5\linewidth]{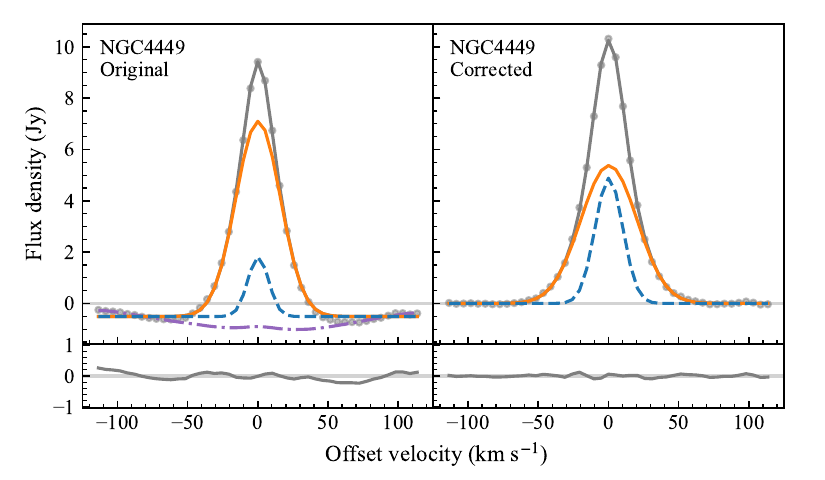}}
\subfloat[]{\includegraphics[width=0.5\linewidth]{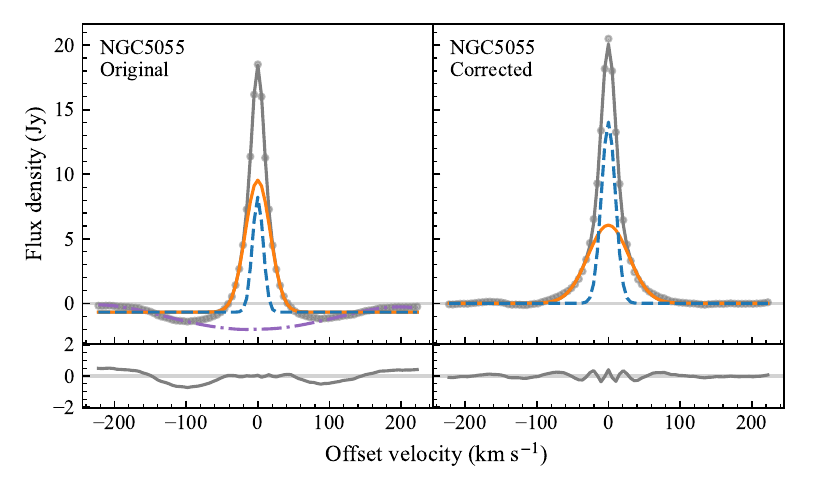}}\\
\subfloat[]{\includegraphics[width=0.5\linewidth]{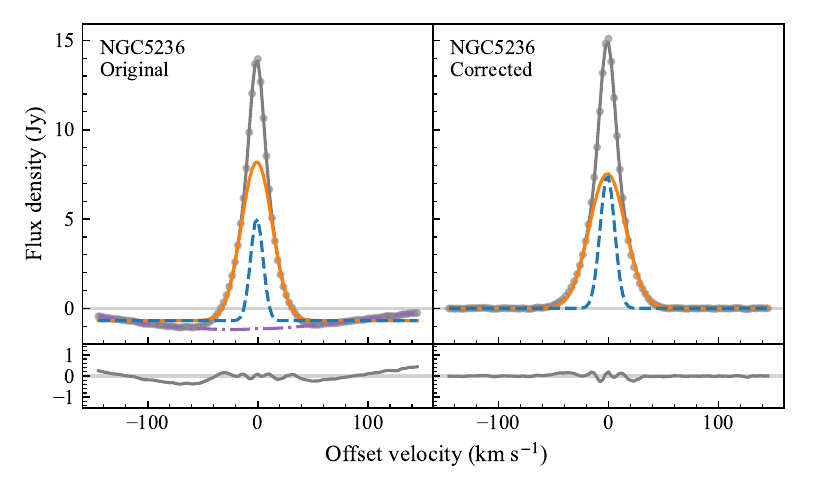}}
\subfloat[]{\includegraphics[width=0.5\linewidth]{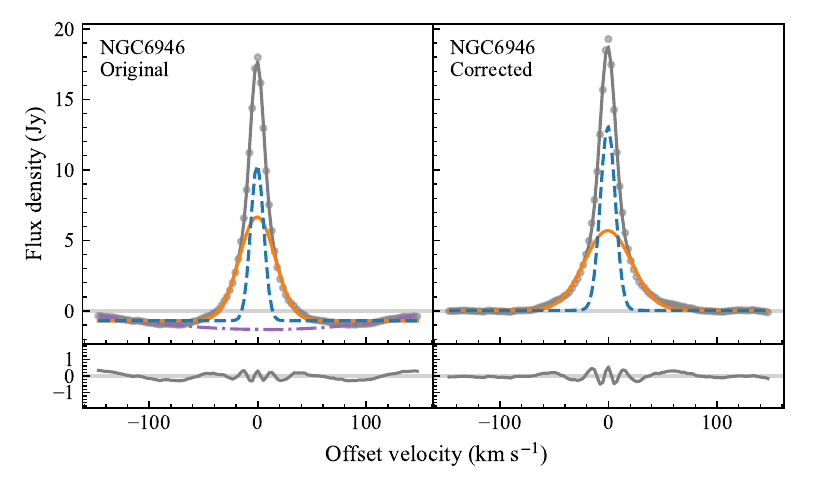}}\\
\caption{The negative-bowl-corrected super-profiles with their {\sc 2gfit} results. Each left panel show the super-profile before correction with their negative bowl fitted with polynomial, and right panel show the corrected super-profile. The purple dash-dotted line in each left panel show the fitted polynomial on the negative bowls.}
\label{fig:GFIT_nbcorr}
\end{figure*}

\begin{figure*}[!p]
\centering
\includegraphics[width=\linewidth]{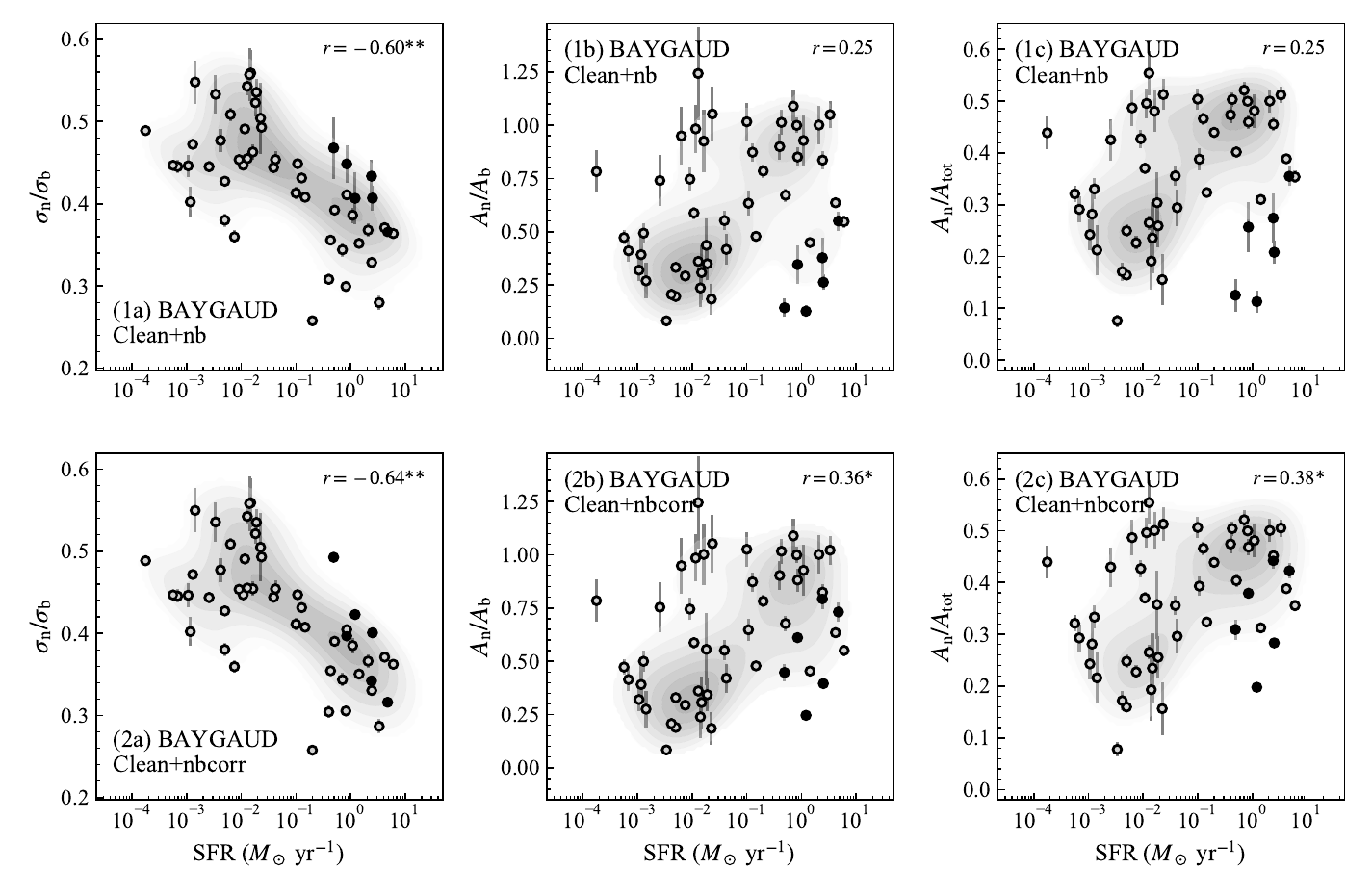}
\caption{Correlation of clean sample with negative-bowl-corrected sample. Upper row shows correlation using super-profiles with negative bowls that are not corrected. Bottom row shows correlation with negative-bowl-corrected sample. The super-profiles with negative bowls are marked with filled circles.\label{fig:correlation_nbcorr}}
\end{figure*}

\endgroup
\twocolumn

\end{document}